%% file: main.tex
\documentclass[10pt,conference]{IEEEtran}
\IEEEoverridecommandlockouts
\usepackage{cite}
\usepackage{amsmath,amssymb,amsfonts}
\usepackage{algorithmic}
\usepackage{graphicx}
\usepackage{textcomp}
\usepackage{xcolor}
\usepackage{booktabs} 
\usepackage{inputx}
\usepackage{dsfont}
\usepackage{tikz}
\usepackage{pgfplots}
\usepackage{multirow}
\usepackage{hyperref}
\def\BibTeX{{\rm B\kern-.05em{\sc i\kern-.025em b}\kern-.08em
    T\kern-.1667em\lower.7ex\hbox{E}\kern-.125emX}}
\newtheorem{definition}{Definition}

\begin{document}

\title{Optimal by Design: Model-Driven Synthesis of Adaptation Strategies for Autonomous Systems
}

\author{\IEEEauthorblockN{Yehia Elrakaiby}
\textit{University of Luxembourg}\\
Luxembourg, Luxembourg \\
yehia.elrakaiby@uni.lu
\and
\IEEEauthorblockN{Paola Spoletini}
\textit{Kennesaw State University}\\
Georgia, USA \\
pspoleti@kennesaw.edu
\and
\IEEEauthorblockN{Bashar Nuseibeh}
\textit{The Open University}\\
Milton Keynes, UK\\
b.nuseibeh@open.ac.uk}

\graphicspath{{Img/}}
\newcommand{\reactt}{\textsc{ObD}}
\newcommand{\react}{\textsc{ObD }}
\newcommand{\yes}{$\checkmark$}
\newcommand{\no}{$\times$}
\newcommand{\ps}{$\circ$}
\newcommand{\na}{$-$}
\newcommand{\br}[1]{\langle #1 \rangle}
\newcommand{\comments}[1]{}

\newcommand{\npaola}[1]{{{\textcolor{red}{ Paola: }}}{\textbf{\textcolor{red}{\textsl{#1}}}}}
\newcommand{\paola}[1]{}

\newcommand{\yehia}[1]{{{\textcolor{blue}{ Yehia: }}}{\textbf{\textcolor{green}{\textsl{#1}}}}}

\newcommand{\bashar}[1]{{{\textcolor{blue}{ Bashar: }}}{\textbf{\textcolor{green}{\textsl{#1}}}}}

\maketitle

\begin{abstract}
	Many software systems have become too large and complex to be managed efficiently by human administrators, particularly when they operate in {\it uncertain} and {\it dynamic} environments and require frequent changes. 
Requirements-driven adaptation techniques have been proposed to endow systems with the necessary means to {\it autonomously} decide ways to satisfy their requirements. However, many current approaches rely on general-purpose languages, models and/or frameworks to design, develop and analyze autonomous systems. Unfortunately, these tools are not tailored towards the characteristics of adaptation problems in autonomous systems. In this paper, we present \textsc{Optimal by Design} (\react), a framework for model-based requirements-driven synthesis of {\it optimal adaptation strategies} for autonomous systems. \react~proposes a model (and a language) for the high-level description of the basic elements of self-adaptive systems, namely the system, capabilities, requirements and environment. Based on those elements, a Markov Decision Process (MDP) is constructed to compute the {\it optimal} strategy or the most rewarding {\it system behavior}. Furthermore, this defines a {\it reflex} controller that can ensure {\it timely} responses to changes. One novel feature of the framework is that it benefits both from goal-oriented techniques, developed for requirement elicitation, refinement and analysis, and {\it synthesis} capabilities and extensive research around MDPs, their extensions and tools. Our preliminary evaluation results demonstrate the practicality and advantages of the framework.
\end{abstract}

\begin{IEEEkeywords}
Autonomous Systems, Markov Decision Process, Controller Synthesis, Optimal Strategies, Adaptive Systems, Requirements Engineering,  Model-driven Engineering, Domain Modeling Language
\end{IEEEkeywords}

	\input{Introduction.tex}

\input{Context}

\input{ReactOverview}

\input{Overview}
\input{Syntax}

\input{SemanticsInprogress}

\input{Evaluation}



\input{Limitations.tex}

\input{RelatedWork.tex}
\input{Conclusion.tex}

\section*{Acknowledgment}
This work was supported, in part, by Science Foundation Ireland grant 13/RC/2094 and ERC Advanced Grant 291652.

\bibliographystyle{IEEEtran}
\bibliography{IEEEabrv,main.bib}
\clearpage 
\newpage 
\input{AppendixB}

\end{document}

%% file: Introduction.tex
\section{Introduction}
\label{sec:introduction}

Autonomous systems such as unmanned vehicles and robots play an increasingly relevant role in our societies. 
Many factors contribute to the complexity in the design and development of those systems. First, they typically operate in dynamic and uncontrollable environments~\cite{Salehie2009,Cheng2009a,Lemos2011,Esfahani2013,Moreno2015}. Therefore, they must continuously adapt their configuration in response to changes, both in their operating environment and in themselves. Since the frequency of change cannot be controlled, decision-making must be almost instantaneous to ensure timely responses. 
From a design and management perspective, it is desirable to minimize the effort needed to design the system and to enable its runtime updating and maintenance. 

A promising technique to address those challenges is requirements-driven adaptation that endow systems with the necessary means to {\it autonomously} operate based on their requirements. Requirements are prescriptive statements of intent to be satisfied by cooperation of the agents forming the system~\cite{Lamsweerde2009}. They say what the system will do and not how it will do it~\cite{Zave1997}. Hence, software engineers are relieved from the onerous task of prescribing explicitly how to adapt the system when changes occur. 
Many current requirements-driven adaptation techniques~\cite{Calinescu2010,Filieri2016} 
follow the Monitor-Analyze-Plan-Execute-Knowledge (MAPE-K) paradigm~\cite{Kephart2003}, which usually works as follows~\cite{Weyns2017}: the {\it Monitor} monitors the {\it managed system and its environment}, and updates the content of the {\it Knowledge} element accordingly; the {\it Analyse} activity 
 uses the up-to-date knowledge to determine whether there is a need for adaptation of the managed element according to the adaptation goals that are available in the knowledge element. If adaptation is required, the {\it Plan} activity puts together a plan that consists of one or more adaptation actions. The adaptation plan is then executed by the {\it Execute} phase. 

This approach 
has two main limitations in highly-dynamic operational environments. First, it tends to be myopic since the system adapts in response to changes without anticipating what the subsequent adaptation needs will be~\cite{Moreno2015} and, thus, it does not guarantee the optimality of the {\it overall behavior} of the autonomous system. This is particularly crucial for systems that have to operate continuously without interruption over long periods of time, e.g., cyber-physical systems. Second, the time to plan adaptations could make timely reaction to changes impossible, particularly in fast changing environments. Therefore, an approach that enables an almost instantaneous reactions to changes is needed. 

In this paper, we propose the Optimal by Design (\react) framework as a first step towards dealing with the aforementioned challenges. 
\react supports a model-based approach to simplify the high-level design and description of autonomous systems, their capabilities, requirements and environment. Based on these high-level models, \react constructs a Markov Decision Process (MDP) that can then be solved (possibly using state-of-the-art probabilistic model checkers) to produce optimal {\it strategies} for the autonomous system. These strategies define {\it optimal reflex} controllers that ensure the ability of autonomous systems to behave optimally and almost instantaneously to changes in itself or its environment. 

Several previous works~\cite{Camara2014a,Camara2015,Moreno2015,Camara2016} encode adaptation problems using general-purpose languages such as those proposed by probabilistic model checkers, e.g., PRISM~\cite{Kwiatkowska2011}. Unfortunately, these languages do not offer primitives tailored to the design and analysis of autonomous systems. This makes them unsuitable to adequately describe the software requirements~\cite{Lamsweerde2009} of the autonomous system and the environment in which it operates. Examples of limitations of these languages resolved in this paper through \react are the Markovian assumption~\cite{Bacchus96} and the implicit-event model~\cite{boutilier99}. 

In a nutshell, \react introduces a novel Domain Specific Modeling Language (DSL) for the description of autonomous systems, its environment and requirements. The semantics of the DSL is then defined in terms of a translation into a Markov Decision Process (MDP) model to enable the synthesis of optimal controllers for the autonomous system. This separation between the model (i.e. the DSL) and its underlying computational paradigm (i.e. MDP), brings several important advantages. First, the level of abstraction at which systems have to be designed is raised, simplifying their modeling by software engineers. Second, requirements become first-class entities, making it possible to elicit them using traditional requirements engineering techniques~\cite{Lamsweerde2009,Sawyer2010,Whittle2010,Morandini2017} and to benefit from goal refinement, analysis and verification techniques developed for goal modeling languages. Moreover, this approach clarifies the limitations of the underlying computational model, namely the aforementioned Markovian assumption and the implicit-event model, and permits the identification and implementation of extensions necessary to overcome those limitations and support the required analysis, verification and reasoning tasks.

The remainder of this paper is structured as follows. Section~\ref{sec:running_example} presents a motivating example, which will be used as a running example throughout the paper. Section~\ref{sec:framework_overview} presents an overview of the \react framework. Section~\ref{sec:model_language} introduces the framework's model and language. Section~\ref{sec:controller_synthesis} provides the semantics of \react models by presenting their translation into MDPs. Section~\ref{sec:evaluation}presents an evaluation of the framework. Section~\ref{sec:limitations} discusses limitations and threats to validity. Finally, Section~\ref{sec:related_work} discusses related work and Section~\ref{sec:conclusion} concludes the paper and presents future work.

%% file: Context.tex
\section{Motivating Example}
\label{sec:running_example}
Our running example, inspired by one of the examples in~\cite{Skubch2013}, is the restaurant $FoodX$. Serving at $FoodX$ is $RoboX$, an autonomous mobile robot. The restaurant comprises three separate sections: (1) the kitchen, (2) the dining area and (3) the office. $RoboX$ is equipped with various sensors to monitor its environment and actuators to move around the restaurant and perform different tasks. 

Several challenges must be dealt with in order to develop a controller for $RoboX$. First, there are events that occur in the environment beyond $RoboX$'s control. For example, a client may request to order or a weak battery signal may be detected. There is also the uncertainty in action effects caused by imperfect actuators, e.g. moving to the kitchen from the dining room could sometimes fail, possibly due to the movement of customers in the restaurant. $RoboX$ may also have multiple (possibly conflicting) requirements: it may have to serve customers' food while it is still hot but also has to keep its batteries charged at all times. Thus, $RoboX$ should be able to prioritize the satisfaction of its requirements, taking into account the effects of their satisfaction over the long-term. It is also desirable that $RoboX$ acts proactively. For instance, waiting in the dining area should be preferred to staying in, for example, the kitchen if doing so would increase the likelihood of it getting orders from customers. 

Since the time and frequency of change in the environment cannot be controlled, enabling immediate and optimal responses to changes is highly desirable. In reactive approaches, {\it classical planning} (e.g., STRIPS~\cite{Fikes1971} and PDDL~\cite{mcdermott98pddl} planners) is often used to determine the best course of action after detection of change. This approach has important limitations. For example, imagine a situation where, while $RoboX$ is moving to serve a customer in the dining room, a weak battery signal is detected. In this case, $RoboX$ can either halt the execution of the current plan until a new plan is computed or continue pursuing serving the customer, having no guarantees that this plan is still the optimal course of action. If the frequency of changes in the environment is high, then the autonomous system may get permanently stuck computing new plans, or be always following sub-optimal plans. 

This example highlights the five requirements for the software to control $RoboX$ which we explore in this paper: 


\begin{enumerate}
	\item Handling of uncertainty in event occurrences and effects;
	\item Proactive and long-term behavior optimization to consider the possible evolution of the system when determining the best course of action;
	\item Fast and optimal response to changes to ensure their ability to operate in highly dynamic environments;
	\item Support of requirements trade-offs and prioritisation;
	\item Support of requirements-driven adaptation to raise the level of abstraction of system design.
\end{enumerate}

\comments{
\paragraph{{\bf C$_1$} - Handling of uncertainty} in occurrence of events and action effectsMany sources of uncertainty have been identified in self-adaptive systems~\cite{Esfahani2013}. In our running example, two main sources of uncertainties can be identified. Firstly, there are exogenous events that occur in the environment beyond $RoboX$'s control. 

\paragraph{Proactive and long-term behavior optimization} As opposed to reactive adaptation where the system only reacts when a requirement is violated, proactive behavior considers the possible evolution of the system to determine the best course of action. 

\paragraph{Fast and optimal response to change}\paola{this is a fuzzy quantity} In an uncontrollable environment, the time and frequency of changes cannot be controlled. Guaranteeing immediate and optimal responses to changes is therefore highly desirable. Traditionally, adaptation is performed using a  Monitor Analyze Planning Execute (MAPE) paradigm where classical planning (see for example STRIPS~\cite{Fikes1971} PDDL~\cite{mcdermott98pddl}) is used to determine the best course of action when a change is detected. This reactive paradigm is inadequate, unless the computation of plans is instantaneous. For example, imagine if while $RoboX$ is moving to serve a customer in the dining room, a weak battery signal is detected. In this case, it is possible to halt the execution of the plan until a new plan is computed or resume the execution of the current plan with no guarantees of its optimality. If the frequency of changes detected in the environment is high, then this situation is worsened as the autonomous system would be permanently stuck in a state where it is computing plans without acting, or it would be indefinitely pursuing plans with no guarantees of their optimality. The computation of an {\it optimal strategy} provides a solution to this problem. An optimal strategy identifies the best adaptation action that should be taken in every possible evolution of the system. Thus, it enables an almost instantaneous response to every possible change since optimal responses are precomputed, eliminating the need to re-plan after every change\paola{I don’t think it is clear how this guarantee instantaneous respone }. In other words, the computation of a strategy defines a {\it reflex controller} for the autonomous system that enables it to optimally and instantaneously react to every possible future change.

 \paragraph{Support of a model-based-requirements-driven adaptation}\paola{Why the kind of the adaptation is part of the requirement? Shouldn’t it be part of the solution?} Requirements are prescriptive statements of intent to be satisfied by cooperation of the agents forming the system~\cite{Lamsweerde2009}. They say what the system will do and not how it will do it~\cite{Zave1997}.\paola{This shouldn’t be part of this but outside} In complex and nondeterministic environments, it is very difficult to anticipate and explicitly prescribe the system's ideal behavior in every situation. A model-based approach where requirements are used to drive system adaptation provides several advantages. First, it raises the level of abstraction at which the system should be designed and analyzed and allows to benefit from works on requirements elicitation, refinements and analysis produced over the years in Requirements Engineering (RE)~\cite{Lamsweerde2009,Sawyer2010,Whittle2010,Morandini2017}\paola{why?}. Second, it relieves the software engineer from the heavy task of prescribing the strategies that are needed to adapt the system when changes occur.
}

\comments{
\subsection{Summary of Challenges}
\label{sec:challenges}

The software of an autonomous robot such as $RoboX$ must satisfy various requirements. First, it must 

enable it to satisfy its various goals and requirements. For example, after getting an order, $RoboX$ must also ensure that the order is given to the kitchen and that subsequently food is delivered to the table. It may also need to charge its batteries after a low battery signal is detected. Obviously, these requirements may conflict. Therefore, $RoboX$ must be endowed with the intelligence necessary to optimize the satisfaction of its various (possibly conflicting) requirements according to the present situation.

There are several ways to define a controller for $RoboX$. One possibility is to predefine all possible situations and specify how the robot should behave in each. This solution is impractical as there are potentially millions of situations which would need to be covered. To rely on the Monitor Analyze Planning Execute (MAPE-K) paradigm could be a better alternative: the robot monitors the environment, plans its best course of action accordingly and then executes the plan. However, the use of planning has some shortcomings. For example, it is not suitable for handling of uncertainty during plan execution. For example, if while the robot is executing some plan, a customer places an order, should the current plan be abandoned or resumed? If the robot decides to resume the plan, then it might be pursuing a sub-ideal plan. On the other hand, if it decides to recompute its plan, then frequent occurrences of events in its environment may prevent it from executing any plan altogether. Furthermore, actuators of $RoboX$ may not be perfect and may not necessary always produce their desired effects. For example, it could be the case that moving to the kitchen from the dining room may sometimes fail, due to the movement of customers in the restaurant. 

Besides dealing with uncertainty, there is the need to deal with multiple (possibly conflicting) requirements. For example, this may happen if $RoboX$ should charge its batteries and bring food to a customer's table while it is still hot. There is therefore a need to support disciplined trade-off of requirements that is based on reasoning about the importance of requirement satisfaction. In order to do so, there should be a reward or utility function which specifies the expected return or payoff for requirements satisfaction and $RoboX$ should choose its actions such that they maximize  this reward or utility function. 

Thus, to deal with the aforementioned challenges, the software of $RoboX$ should aim at the {\it optimal} satisfaction of its {\it requirements}~\cite{Zave1997,Lamsweerde2009} given a description consisting of the system and its environment. Notice that requirements say what the system will do and not how it will do it~\cite{Zave1997}. Thus, they enable to raise the level of abstraction of system specifications and reduce considerably the size of system specifications that have to be provided. 

To identify the optimal behavior of the system based on a high-level description of the system, its environment and its requirements, we construct a Markov Decision Process (MDP). By solving this MDP, we can compute an {\it optimal} strategy that identifies the action that should be taken in every possible situation in order to best satisfy the requirements of $ROboX$. 

One important benefit of this approach is that relying on requirements for the generation of strategies makes possible to benefit from works on requirements elicitation, refinements and analysis produced over the years in Requirements Engineering (RE)~\cite{Lamsweerde2009,Sawyer2010,Whittle2010,Morandini2017}. It also enables to benefit from the powerful computational machinery that comes with Markov Decision Models and its extensions and tools.

We summarize the following main challenges for the software that controls $RoboX$:
\begin{itemize}
	\item High-level specification of goals/requirements: it is impractical to predict every possible situation and predefine a good response to it. It would therefore be much more convenient to support the specification of high-level goals for the robot and provide it with the means necessary to determine its best course of action according to the current context or situation;
 
	\item Handling uncertainty in the environment: $RoboX$ must be able to react optimally, and instantly if possible, even when exogenous events beyond its control occur in its environment;

	\item Handling of non-determinism: actuators of $RoboX$ may not always produce the desired outcomes. Similarly, events may not have deterministic effects. Therefore, it is highly desirable that this kind of uncertainty be considered when $RoboX$ decides of its best course of action.
\end{itemize}


}

%% file: ReactOverview.tex
\section{Framework Overview}
\label{sec:framework_overview}
\react is a framework for the model-based requirements-driven synthesis of optimal adaptation strategies for autonomous systems. The model-based approach raises the level of abstraction at which systems need to be described and simplifies model maintenance and update. Adaptation in \react is {\it requirements-driven}, enabling systems to autonomously determine the best way to pursue their objectives. 
Based on \react models, Markov Decision Processes (MDPs) are constructed. Solving those MDPs determines the system's optimal strategy, i.e., the  behavior that maximizes the satisfaction of requirements. In a strategy, the best adaptation action that should be taken in every possible (anticipated) future evolution of the system is identified, eliminating the need to re-analyze and re-plan after every change and enabling almost instantaneous reactions. Indeed, an optimal strategy defines a {\it reflex controller} that can react optimally and in a timely way.

\begin{figure}
	\centering
	\includegraphics[width=0.92\linewidth]{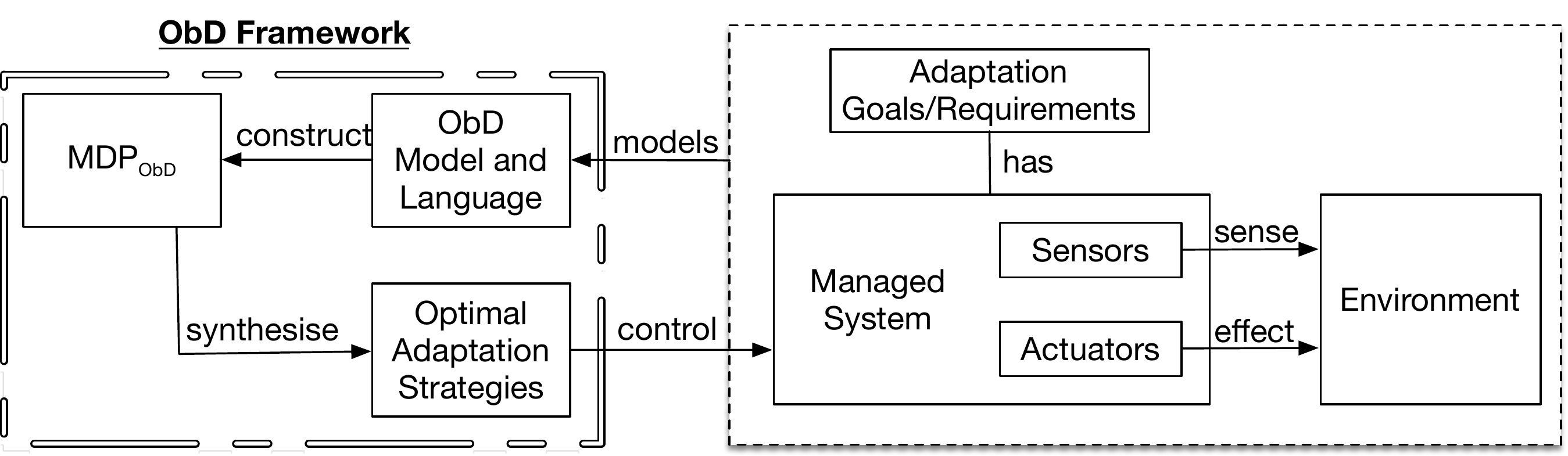}
	\caption[Restaurant Layout]{Framework Overview}
	\label{fig:react_overview}
\end{figure}

Figure~\ref{fig:react_overview} depicts an overview of the framework, which includes a model (and a language) to describe the basic elements of self-adaptive systems~\cite{Weyns2017}: the {\it environment} refers to the part of the external world with which the system interacts and in which the effects of the system will be observed and evaluated; the {\it requirements} or adaptation goals are the concerns that need to be satisfied; the {\it managed system} represents the application code or capabilities/actuators that can be leveraged to satisfy the requirements. Based on these elements, the {\it controller} or the managing system ensures that the adaptation goals are satisfied in the managed system, is synthesized.

\comments{
From an engineering perspective, \react can be integrated into the architecture proposed by Kramer and Magee's for the engineering of self-adaptive software systems [31]. Their architecture comprises of three layers~\cite{Weyns2017}: (1) the bottom layer (the component layer) consists of the interconnected components that provide the functionalities of the system, (2) a middle {\it change management} layer which consists of a set of pre-specified plans. The middle layer reacts to status changes of bottom layer by executing plans with change actions that adapt the component configuration of the bottom layer. The middle layer is also responsible for effecting changes to the underlying managed system in response to new objectives introduced from the layer above. (3) The top layer, {\it Goal Management}, comprises a specification of high-level goals. This layer produces change management plans in response to requests for plans from the layer beneath or after new goals are introduced by stakeholders. REact can be integrated in this architecture as follows: the {\it goal management} layer would comprise REact domain models which would be translated into MDPs. Solving those MDPs would generate the {\it change management} layer or the plans for reacting to changes at the bottom layer. A request for a new set of plans, or in other words the computation a new strategy, would be produced by the change management layer whenever there is a need to update the models of the component layer or of the environment, or after the introduction of new system goals or requirements by stakeholders.

One important benefit of this approach is that relying on requirements for the generation of strategies makes possible to benefit from works on requirements elicitation, refinements and analysis produced over the years in Requirements Engineering (RE)~\cite{Lamsweerde2009,Sawyer2010,Whittle2010,Morandini2017}. It also enables to benefit from the powerful computational machinery that comes with Markov Decision Models and its extensions and tools.

The software of an autonomous robot such as $RoboX$ must satisfy various requirements. First, it must 

enable it to satisfy its various goals and requirements. For example, after getting an order, $RoboX$ must also ensure that the order is given to the kitchen and that subsequently food is delivered to the table. It may also need to charge its batteries after a low battery signal is detected. Obviously, these requirements may conflict. Therefore, $RoboX$ must be endowed with the intelligence necessary to optimize the satisfaction of its various (possibly conflicting) requirements according to the present situation.

There are several ways to define a controller for $RoboX$. One possibility is to predefine all possible situations and specify how the robot should behave in each. This solution is impractical as there are potentially millions of situations which would need to be covered. To rely on the Monitor Analyze Planning Execute (MAPE-K) paradigm could be a better alternative: the robot monitors the environment, plans its best course of action accordingly and then executes the plan. However, the use of planning has some shortcomings. For example, it is not suitable for handling of uncertainty during plan execution. For example, if while the robot is executing some plan, a customer places an order, should the current plan be abandoned or resumed? If the robot decides to resume the plan, then it might be pursuing a sub-ideal plan. On the other hand, if it decides to recompute its plan, then frequent occurrences of events in its environment may prevent it from executing any plan altogether. Furthermore, actuators of $RoboX$ may not be perfect and may not necessary always produce their desired effects. For example, it could be the case that moving to the kitchen from the dining room may sometimes fail, due to the movement of customers in the restaurant. 

Besides dealing with uncertainty, there is the need to deal with multiple (possibly conflicting) requirements. For example, this may happen if $RoboX$ should charge its batteries and bring food to a customer's table while it is still hot. There is therefore a need to support disciplined trade-off of requirements that is based on reasoning about the importance of requirement satisfaction. In order to do so, there should be a reward or utility function which specifies the expected return or payoff for requirements satisfaction and $RoboX$ should choose its actions such that they maximize  this reward or utility function. 

Thus, to deal with the aforementioned challenges, the software of $RoboX$ should aim at the {\it optimal} satisfaction of its {\it requirements}~\cite{Zave1997,Lamsweerde2009} given a description consisting of the system and its environment. Notice that requirements say what the system will do and not how it will do it~\cite{Zave1997}. Thus, they enable to raise the level of abstraction of system specifications and reduce considerably the size of system specifications that have to be provided. 

To identify the optimal behavior of the system based on a high-level description of the system, its environment and its requirements, we construct a Markov Decision Process (MDP). By solving this MDP, we can compute an {\it optimal} strategy that identifies the action that should be taken in every possible situation in order to best satisfy the requirements of $ROboX$.

}

%% file: Overview.tex
\section{\react Modeling Language}
\label{sec:model_language}

%% file: Syntax.tex
The computation of {\it optimal strategies} is based on a {\it domain model}. A domain model specifies the environment, the capabilities of the autonomous system (or agent) and its requirements. Formally, an \react model ($\mathcal{D}_r$) is a tuple $\langle \mathcal{SV}, \mathcal{AD}, \mathcal{ED}, \mathcal{RQ}, s_c \rangle$ where: 
\begin{itemize}
	\item $\mathcal{SV}$ is a finite set of state variables with finite domains. State variables describe the possible states, i.e., the configuration of the software system and the environment;
	\item $\mathcal{AD}$ is a finite set of action descriptions representing the means that are available to the agent to change the system state, i.e. update the state variables $\mathcal{SV}$;
	\item $\mathcal{ED}$ is a finite set of event descriptions to represent the uncontrollable occurrences in the environment, i.e., events that change the state beyond the agent's control.
	\item $\mathcal{RQ}$ is a finite set of requirements, i.e., the (operationalisable) goals that the software system should satisfy;
	\item $s_c$ is the initial state of the system 
	determined by the agent's monitoring components and sensors.
\end{itemize}
 An \react model has a corresponding textual representation called its {\it domain description}. It is formalized in the following using a variant of Backus-Naur Form (BNF): the names enclosed in angular brackets identify non-terminals, names in bold or enclosed within quotation marks are terminals, optional items are enclosed in square brackets,~$|$ is "or", items repeated one or more times are suffixed with $+$ and parentheses are used to group items together.

\subsection{State, Actions and Events} 
\begin{figure*}[t]
	\centering
	\includegraphics[width=0.8\linewidth]{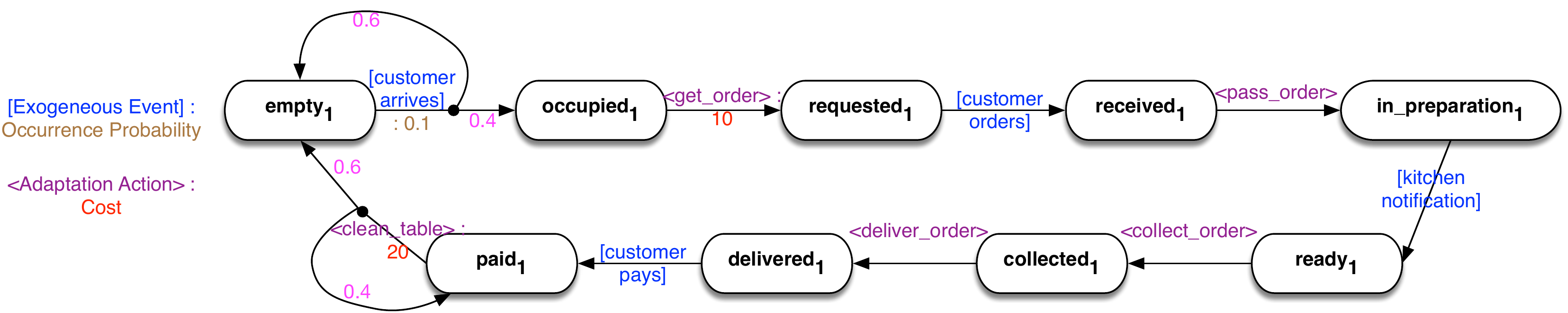}
	\caption{A simplified model of serving a table at restaurant $FoodX$.}
	\label{fig:tablelts}
\end{figure*}

\paragraph*{State Variables ($\mathcal{SV}$)} define the possible {\it states}, i.e., configurations of the software system and the environment. A variable $x\in \mathcal{X}_s$ is a multi-valued variable with a corresponding domain, denoted $dom(x)$. Every value $y\in dom(x)$ is a configuration of $x$. A state variable is defined as follows:
 \begingroup\makeatletter\def\f@size{9}\check@mathfonts
\def\maketag@@@#1{\hbox{\m@th\large\normalfont#1}} 
\begin{align*}
&\br{SV} ::= \textbf{Variable}\textbf{ } \br{ID} \, \textbf{domain} \,  ``\{" \, \br{VALS} \, ``\}"\\
&\br{VALS} ::= \br{ID} \, | \, \br{ID} \, ``," \, \br{VALS}  
\end{align*} \endgroup

\noindent where $\br{ID}$ is {\it text}, i.e., a concatenation of letters, digits and symbols. For example, we can represent the location of $RoboX$ and the status of tables at the restaurant using the following variables:
\begingroup\makeatletter\def\f@size{9}\check@mathfonts
\def\maketag@@@#1{\hbox{\m@th\large\normalfont#1}}%
\begin{align*}
&\textbf{Variable } location \textbf{ domain } \{atTable_1,atTable_2,atTable_3,\\
& \hspace{0.5cm}\hspace{0.5cm} atTable_4,inDining\_room,inKitchen,inOffice\}\\
&\textbf{Variable } table_i \hspace{1cm}(\forall \, 1\leq i \leq 4)\\
& \hspace{0.5cm}\textbf{ domain } \{empty,occupied,requested,received,\\
& \hspace{0.5cm}\hspace{0.5cm}in\_preparation,ready,collected,delivered,paid\}
\end{align*}
\endgroup
\noindent The variable $location$ defines the possible locations of $RoboX$. The variables $table_i$ represent the status of tables: when there are no customers at $table_i$, then $table_i\text{=}empty$. When a customer arrives and sits at the table, $table_i$ becomes $occupied$. Figure~\ref{fig:tablelts} depicts the update of the value of $table_i$ with the occurrence of the robot actions \{{\it get\_order, give\_order, collect\_order, deliver\_ order, clean\_table}\} and the exogenous events \{{\it customer\_arrives, customer\_orders, kitchen\_notification, custo\-mer\_pays}\}. In contrast to actions, exogenous events have an occurrence probability denoting their likelihood in a given situation. Actions, on the other hand, have a cost that represent the effort or price of their execution. Both exogenous events and actions do not have to be deterministic, i.e., their execution can have various effects, each with a different probability (in pink in the figure).

\comments{When $RoboX$ asks the customer to order, $table_i\text{=}requested$. When the customer places an order, $table_i\text{=}order\_received$. After $RoboX$ gives the order to the kitchen, $table_i\text{=}in\_preparation$. After the kitchen notifies $RoboX$ that food is ready, $table_i\text{=}ready$. When $RoboX$ collects the food from the kitchen, $table_i\text{=}collected$. After food is delivered to the table, $table_i\text{=}delivered$. After the customer pays the bill and leaves, $table_i\text{=}paid$. When $RoboX$ cleans the table, $table_i\text{=}empty$.
}
Variables which are not explicitly defined are considered to be boolean, i.e., their domain is $\{tt,\textit{ff}\}$. The notations $id$ and $!id$ are used as shortcuts for $id\text{=}tt$ and $id\text{=}\textsl{ff}$, respectively. 
The following declaration defines a boolean variable to represent that customers sitting at a table had looked at the menu.
 \begingroup\makeatletter\def\f@size{9}\check@mathfonts
\def\maketag@@@#1{\hbox{\m@th\large\normalfont#1}} 
\begin{align*}
&\textbf{Variable } looked_i &(\forall \, 1\leq i \leq 4)
\end{align*}
\endgroup

\paragraph*{Actions ($\mathcal{AD}$)} are means that are available to the agent to change the system state. An action description is an expression $\br{AD}$ that is defined as follows:
 \begingroup\makeatletter\def\f@size{9}\check@mathfonts
\def\maketag@@@#1{\hbox{\m@th\large\normalfont#1}} 
\begin{align*}
&\br{AD} ::= \textbf{Action} \, \br{ID} \, \br{PEFFS}^+ \, [\textbf{cost} \, \br{\mathds{N}}]\\
&\br{PEFFS} ::= \textbf{ if } \br{CND} \textbf{ effects }  \br{EFFS}^+ \\
&\br{EFFS} ::= ``\langle" \br{EFF}^+ \,[ \textbf{prob } \br{P}] ``\rangle"\\
&\br{CND} ::= \br{ATOM} \,| \, \br{BL} \,| \,``!" \br{CND}\, | \,\br{CND} \\
& \hspace{0.5cm}  \, ``{\bf \&}" \, \br{CND} \,| \, \br{CND} \, ``{\bf ||}" \, \br{CND}  \,| \, ``(" \br{CND} ``)"\nonumber\\
&\br{EFF} ::=  \br{ID} ``{\bf =}" \br{ID}\\
&\br{ATOM} ::= \br{ID} \, | \, ``!"\br{ID} \,| \, \br{ID} ``{\bf =}" \br{ID}\\
&\br{BL} ::= ``true" \, | \, ``false"
\end{align*} \endgroup

\noindent 
Actions can have a cost representing the difficulty level or effort necessary to execute it. Action costs are useful to trade-off the satisfaction of requirements with the required effort and, when not specified, are set to zero. 
In the following example, the cost of moving to $table_i$ is set to $10$.
\comments{
\begingroup\makeatletter\def\f@size{9}\check@mathfonts
\def\maketag@@@#1{\hbox{\m@th\large\normalfont#1}}
&\textbf{Action } move\_to\_table_i&(\forall \, 1\leq i \leq 4)\\
& \hspace{1cm} \textbf{ if } location=inDining\_room \textbf{ effects }\langle location=atTable_i \rangle \\
& \hspace{1cm}\textbf{ cost } 10
\end{align*}
\endgroup
While in the above example the effect associated to the action is only one and so it will happen with probability $1$, in general
the effect of an action is probabilistic, i.e., different effects correspond to an action and each effect is associated with a certain probability to occur,
For example, it is possible to specify that the action to move to the kitchen from the dining room has the desired effect with probability and does not bring any change with probability $20\%$: 
}
\begingroup\makeatletter\def\f@size{9}\check@mathfonts
\def\maketag@@@#1{\hbox{\m@th\large\normalfont#1}} 
\begin{align*}
&\textbf{Action } move\_to\_kitchen \textbf{ if } location\text{=}inDining\_room\\
&\hspace{0.5cm} \textbf{ effects } \langle location\text{=}inKitchen \textbf{ prob } 0.8 \rangle\\
&\hspace{1cm} \langle location\text{=}inDining\_room \textbf{ prob } 0.2 \rangle \hspace{0.5cm} \textbf{ cost } 10
\end{align*}
\endgroup

Note that an expression $\br{AD}$ is well-formed only if (1) its various $\br{CND}$ are disjoint, i.e., they cannot be satisfied at the same time and (2) for every $\br{PEFFS}$, the sum of the probabilities $\br{P}$ of its subexpressions $\br{EFFS}$ is one, i.e., $\sum_{i=1}^{|\br{EFFS}|}\br{P}=1$. Note that we allow $\sum_{i=1}^{|\br{EFFS}|}\br{P}$ to be less than one. In this case, action execution has no effect with a probability of $1-\sum_{i=1}^{|\br{EFFS}|}\br{P}$. For example, this makes it possible to remove the second effect, $\langle location=inDining\_room \textbf{ prob } 0.2 \rangle$, from the previous action description without affecting the action semantics.


\paragraph*{Events ($\mathcal{EV}$)} represent occurrences that are not controlled by the agent. They may happen in the environment at any moment. An event description is expressed as follows:
 \begingroup\makeatletter\def\f@size{9}\check@mathfonts
\def\maketag@@@#1{\hbox{\m@th\large\normalfont#1}} 
\begin{align*}
&\br{EV} ::= \textbf{Event} \, \br{ID} \, \br{PEFFS}^+ \\
&\br{PEFFS} ::= \textbf{ if } \br{CND} \, [\textbf{occur prob}\, \br{P} ]  \, \textbf{ effects } \br{EFFS}^+ 
\end{align*} 
\endgroup

\comments{
For example, it is possible to model the effects of a customer request to order using an {\it exogenous} event description as follows:
\begingroup\makeatletter\def\f@size{9}\check@mathfonts
\def\maketag@@@#1{\hbox{\m@th\large\normalfont#1}}
&		\textbf{Event } request\_to\_order_i\hspace{1cm}(\forall \, 1\leq i \leq 4)\\
& \hspace{1cm} \textbf{ if } table_i=occupied \textbf{ effects } table_i=requested
\end{align*}
\endgroup
}
Events are conditional and can occur with a different probability depending on the situation. 
For instance, we can represent that customers may order with a higher probability if they had looked at the menu as follows:
 \begingroup\makeatletter\def\f@size{9}\check@mathfonts
\def\maketag@@@#1{\hbox{\m@th\large\normalfont#1}} 
\begin{align*}
&		\textbf{Event } request\_to\_order_i \hspace{1cm}(\forall \, 1\leq i \leq 4)\\
& \hspace{0.5cm}\textbf{ if } table_i\text{=}occupied \,\&\, looked_i \textbf{ occur prob } 0.9\\
&\hspace{1cm} \textbf{ effects } \langle table_i\text{=}requested \rangle\\
& \hspace{0.5cm}\textbf{ if } table_i\text{=}occupied\, \& \, !looked_i \textbf{ occur prob } 0.2\\
&\hspace{1cm} \textbf{ effects } \langle table_i\text{=}requested \rangle
\end{align*}
\endgroup
\subsection{Requirements}
\label{sec:modeling_requirements} Requirements represent the objectives of the autonomous system. Every requirement is associated with a reward denoting its importance. 
\react currently supports fourteen requirement types, which build upon and extend the goal patterns of the KAOS goal taxonomy~\cite{Letier2002}. Requirements are expressions:
 \begingroup\makeatletter\def\f@size{9}\check@mathfonts
\def\maketag@@@#1{\hbox{\m@th\large\normalfont#1}} 
\begin{align*}
&		\br{RE} ::= \textbf{ ReqID } \br{ID} \, \br{REP}\\
& \br{REP} ::= ((\br{UA} \, | \, \br{UM} \,| \, \br{CA} \, | \, \br{DEA} \, | \, \br{DFA} \, | \, \br{CM} \, |  \\
&\hspace{0.5cm}\, \br{DEM} \, | \, \br{DFM}  \, \br{PM} \, | \, \br{PDEM} \, | \,  \br{PDFM}) \,[\textbf{reward } \mathds{N}]) \,|\, \nonumber\\
& \hspace{0.5cm} ((\br{RPM} \, | \, \br{RPDEM} \, | \, \br{RPDFM}) \,[\textbf{reward\_once } \mathds{N}])\nonumber
\end{align*} \endgroup
\noindent A requirement's type is determined based on whether it: is conditional (C) or unconditional (U); is a {\it maintain} (M) or {\it achieve} (A) requirement: duration of {\it maintain} requirements can be time-limited and its compliance can be best-effort (P) or strict (PR), i.e., during its duration the requirement does not have to be ``always'' satisfied; has a deadline (D), which can be exact (E), i.e., the requirement has to be satisfied at the deadline, or flexible (F), the requirement has to be satisfied within the deadline.
\comments{
\noindent A requirement's type is determined based on whether it:
\begin{itemize}
	\item is conditional (C) or unconditional (U);
	\item is a {\it maintain} (M) or {\it achieve} (A) requirement: duration of {\it maintain} requirements can be time-limited and its compliance can be best-effort (P) or strict (RP), i.e., during its duration the requirement does not have to be ``always'' satisfied;
	\item has a deadline (D), which can be exact (E) or flexible (F), i.e., the requirement has to be satisfied at or within the deadline respectively.
\end{itemize}
}
Due to space limitations, we only present unconditional, conditional and achieve deadline requirements. 


\comments{
\noindent\paragraph*{On Rewards for Maintain Requirements with Duration} In addition to prioritization of requirements, rewards play a second role in the case of maintain requirements which have a fixed duration. In particular, rewards can be given either
\begin{itemize}
	\item at every time step the requirement is complied with,
	\item only if the requirement is complied with throughout its whole duration.
\end{itemize}
\noindent In the former case, {\it partial-compliance} is rewarded, i.e., the agent could envisage to violate the requirement for a period of time if the satisfaction of other requirements is more important and then go back to comply with the requirement again. In the latter case, the agent would have to choose between either {\it full-compliance} or {\it non-compliance} since {\it partial-compliance} is not rewarded. 
}

\textit{Unconditional Requirements:} denote conditions that have to be always maintained or (repeatedly) achieved. 
 \begingroup\makeatletter\def\f@size{9}\check@mathfonts
\def\maketag@@@#1{\hbox{\m@th\large\normalfont#1}} 
\begin{align*}
&  \br{UA} ::= \textbf{achieve} \, \br{CND} && \br{UM} ::= \textbf{maintain} \, \br{CND}
\end{align*} \endgroup

\noindent For example, a $\br{UM}$ requirement to remain in the dining room or an $\br{UA}$ to ensure that $table_1$ repeatedly pays.
 \begingroup\makeatletter\def\f@size{9}\check@mathfonts
\def\maketag@@@#1{\hbox{\m@th\large\normalfont#1}} 
\begin{align*}
&	 \textbf{ReqID } req_1 \textbf{ maintain } location\text{=}inDining\_room \nonumber\\
&	 \textbf{ReqID } req_2 \textbf{ achieve } table_1\text{=}paid \nonumber
\end{align*} \endgroup

\textit{Conditional Requirements:} should be satisfied only after some given conditions are true. They can have a cancellation condition after which their satisfaction is no longer required. 
 \begingroup\makeatletter\def\f@size{9}\check@mathfonts
\def\maketag@@@#1{\hbox{\m@th\large\normalfont#1}} 
\begin{align*}
& \br{CA} ::\text{=} \textbf{achieve} \, \br{CND} \, \textbf{if} \,\br{CND} \,[\textbf{unless} \, \br{CND}] \\
& \br{CM} ::\text{=} \textbf{maintain} \br{CND} \textbf{if} \br{CND} \,[\textbf{unless} \, \br{CND}] 
\end{align*} \endgroup
\noindent For example, $RoboX$ may have to get the order from $table_i$ only if $table_i$ requests to order, or it should remain in the dining room after it gets $table_1$ until $table_1$'s order is served.
\begingroup\makeatletter\def\f@size{9}\check@mathfonts
\def\maketag@@@#1{\hbox{\m@th\large\normalfont#1}} 
\begin{align*}
&		\textbf{ReqID } req_3 \textbf{ achieve } table_i\text{=}received\\
& \hspace{0.5cm} \textbf{ if } table_1=requested \textbf{ reward } 100\\
&		\textbf{ReqID } req_4 \textbf{ maintain } location=inDining\_room \\
& \hspace{0.1cm}\textbf{ if } table_1\text{=}requested \textbf{ unless } table_1\text{=}received \textbf{ reward } 100
\end{align*}
\endgroup

\textit{Deadline Requirements:} must be satisfied after an exact number of time instants or within a period of time:
\begingroup\makeatletter\def\f@size{9}\check@mathfonts
\def\maketag@@@#1{\hbox{\m@th\large\normalfont#1}} 
\begin{align*}
& \br{DEA} ::= \textbf{achieve} \, \br{CND} \,  \textbf{after} \,\mathds{N}_{+} \, \textbf{if} \,\br{CND} \,[\textbf{unless}  \,\br{CND}] \\
& \br{DFA} ::= \textbf{achieve} \, \br{CND} \,  \textbf{within} \,\mathds{N}_{+} \, \textbf{if} \,\br{CND} \,[\textbf{unless} \,\br{CND} ] 
\end{align*}
\endgroup

For example, $RoboX$ may have to be at $table_1$ within at most 4 time units after $table_1$ requests to place an order, or it may have to be at the kitchen after exactly 4 time units after it receives a notification that food is ready.
\begingroup\makeatletter\def\f@size{9}\check@mathfonts
\def\maketag@@@#1{\hbox{\m@th\large\normalfont#1}} 
\begin{align*}
&		\textbf{ReqID } req_5 \textbf{ achieve } location\text{=}atTable_1 \textbf{ within } 4\\
& \hspace{0.5cm} \textbf{ if } table_1\text{=}requested \textbf{ reward } 100\\
&		\textbf{ReqID } req_6 \textbf{ achieve } location\text{=}inKitchen \textbf{ after } 4\\
& \hspace{0.5cm} \textbf{ if } table_1\text{=}ready  \textbf{ reward } 100
\end{align*}
\endgroup

\noindent In the following, we use the terms {\it name}, {\it required condition}, {\it activation condition}, {\it cancellation condition} and {\it deadline} to refer to the parts of a requirement expression that come after $`\text{ReqID'}$, $`\text{achieve'}$ or $`\text{maintain'}$, $`\text{if'}$, $`\text{unless'}$ and $`\text{after'}$ or $`\text{within'}$ parts of the requirement expression respectively. 

%% file: SemanticsInprogress.tex
\section{Controller Synthesis}
\label{sec:controller_synthesis}
Markov Decision Processes (MDPs) are mathematical frameworks for modeling and controlling {\it stochastic dynamical systems}~\cite{boutilier99}. Informally, MDPs may be viewed as Labeled Transition Systems (LTSs) where transitions are probabilistic and can be associated to rewards. Intuitively, solving an MDP means finding an optimal {\it strategy}, i.e., determining the actions to execute in every state in order to maximize the total expected rewards. 
In the following, we first introduce MDPs (Section~\ref{sec:mdp_overview}), then we discuss the main steps needed to construct an MDP starting from an \react domain model (Section~\ref{sec:mdp_construction}).


\subsection{Introduction to MDPs with Rewards}
\label{sec:mdp_overview}
A reward $MDP$ 
is a tuple $\langle \mathcal{S}, \mathcal{A}, \mathcal{T}, \mathcal{R}, \gamma \rangle$, where:
\begin{itemize}
	\item $\mathcal{S}$ is the finite set of all possible states of the system, also called the {\it state space}; 
	\item $\mathcal{A}$ is a finite set of actions;
	
	\item $\mathcal{T}: \mathcal{S}\times \mathcal{A} \times D(\mathcal{S})$ where $D(\mathcal{S})$ is the set of probability distributions over states $S$. A distribution $d(S)\in D(S): S\rightarrow [0,1]$ is a function such that $\Sigma_{s \in S}d(s) = 1$. The transition relation $\mathcal{T}(s_i,a,d)$ specifies the probabilities $d(s_j)$ of going from state $s_i$ after execution of action $a$ to states $s_j$. In the following, we will use the (matrix) notation $Pr_a(s_i,s_j)$ to represent the probability $d(s_j)$ of going to $s_j$ after execution of $a$ in $s_i$;

	\item $\mathcal{R}: \mathcal{S} \times \mathcal{A} \times \mathcal{S} \rightarrow \mathds{R}$ is a reward function specifying a finite numeric reward value $\mathcal{R}(s_i,a,s_j)$ when the system goes from state $s_i$ to state $s_j$ as a result of executing action $a$. Thus, rewards may be viewed as incentives for executing actions. We will use $R_a(s_i,s_j)$ to represent $\mathcal{R}(s_i,a,s_j)$.
	
	
\end{itemize}
Formally, a (memoryless) strategy is a mapping $\pi : \mathcal{S} \rightarrow \mathcal{A}$ from states to actions. An optimal strategy, denoted $\pi^*$, is the one which maximizes the {\it expected linear additive} utility, formally defined as $V^\pi(s)=
\mathds{E}[ \sum_{t'=0}^{\infty} \gamma^{t'} R^{\pi_s}_{t+t'} ]$. 
Intuitively, this utility states that a strategy is as good as the amount of discounted reward it is expected to yield~\cite{Mausam2012}. Setting $\gamma=1$ expresses indifference of the agent to the time in which a particular reward arrives; setting it to a value $0\leq \gamma < 1$ reflects various degrees of preference to rewards earned sooner.\

MDPs have a key property: solving an MDP finds an optimal strategy $\pi^*$, which is deterministic, Markovian and stationary. This means that computed strategies are independent of both past actions/states and time, which ensures their compactness and practicality. Furthermore, there exist practical algorithms for solving MDPs, e.g., value iteration and policy iteration. Both of these algorithms can be shown to perform in polynomial time for fixed $\gamma$~\cite{Littman}. 

\begin{figure*}[!htb]
	\minipage{0.3\textwidth}
	\includegraphics[width=45mm, scale=0.45]{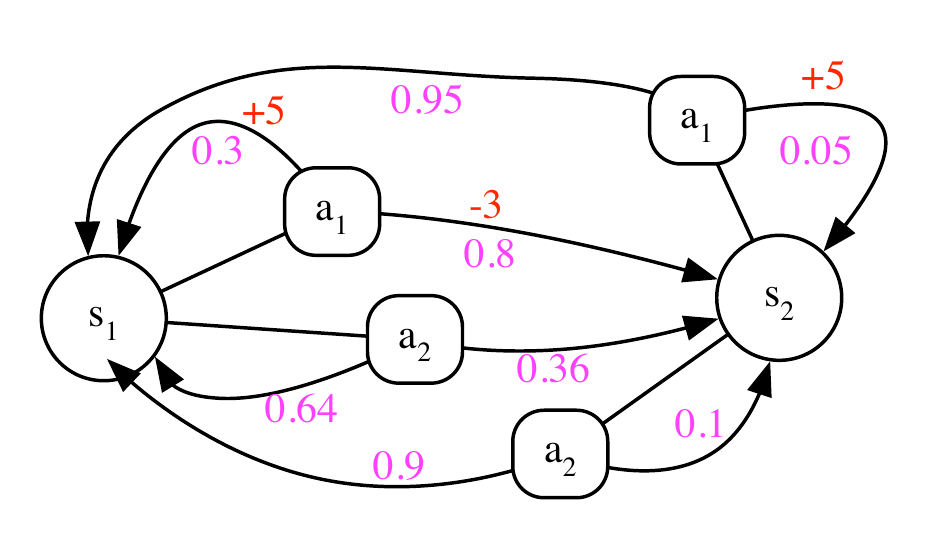}
	\caption{Basic MDP model}\label{fig:mdp_basic}
	\endminipage\hfill
	\minipage{0.35\textwidth}
	\includegraphics[width=55mm, scale=0.5]{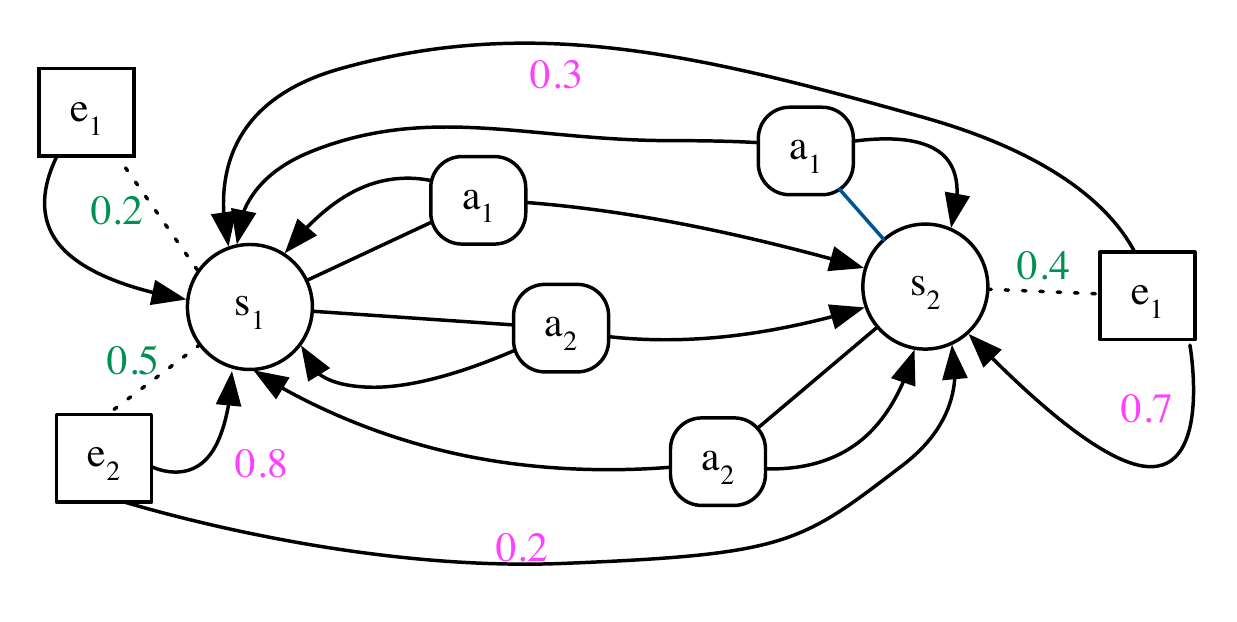}
	\caption{Support of Exogenous Events}\label{fig:mdp_events}
	\endminipage\hfill
	\minipage{0.3\textwidth}
	\includegraphics[width=50mm, scale=0.5]{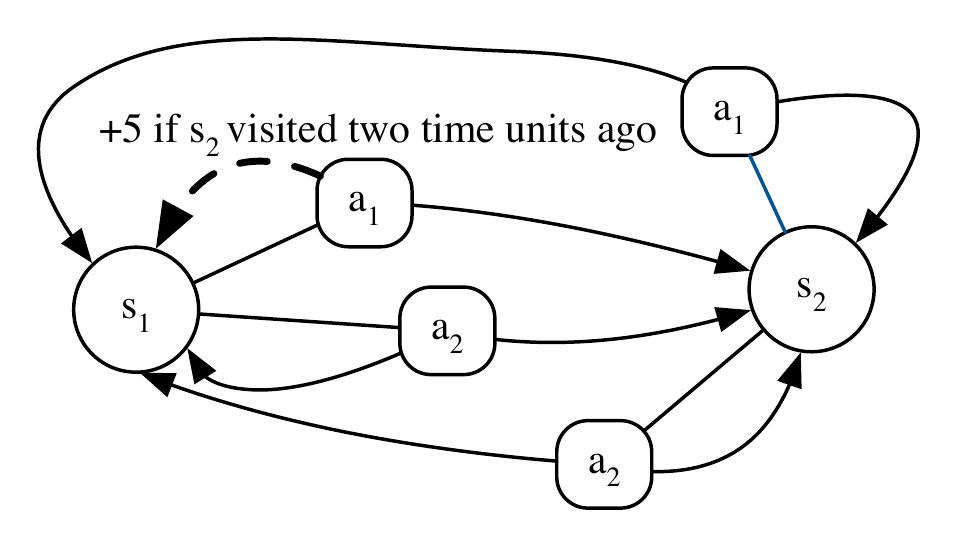}
	\caption{Support of Non-Markovian Rewards}\label{fig:mdp_rewards}
	\endminipage
\end{figure*}

 MDPs with memoryless strategies, depicted in Figure~\ref{fig:mdp_basic} (rounds are states and rounded squares are events), have however the following restrictions: 

	\paragraph*{The {\it implicit-event action model}~\cite{boutilier99}} MDPs do not support an explicit representation of exogenous events.  Figure~\ref{fig:mdp_events} shows exogenous events (in non-rounded squares connected with pointed line to states) that can occur with certain occurrence probabilities (in green in the figure) in every state. Exogenous events are an essential element to model aspects of the environment that are not controllable by the agent. They are the means to represent, for example, that customers can arrive at the restaurant or that they may request to order.
	
	\paragraph*{The Markovian assumption~\cite{Bacchus96,Bacchus97}} in MDP, reward and transition functions have to be Markovian, i.e., they can not refer to the history of previous states or transitions. Figure~\ref{fig:mdp_rewards} shows an example of a non-Markovian reward (described on the dashed transition), i.e., one that is entailed only if certain conditions are satisfied on the history of states and transitions. The support of non-Markovian rewards is necessary to associate transitions that satisfy requirements~\cite{Lamsweerde2009}, which are often conditional and can have deadlines, with rewards. 

\subsection{Overview of the Construction of MDPs from \react Models}
\label{sec:mdp_construction}
The construction of MDPs based on \react models\footnote{The formal details can be found in \url{https://goo.gl/aoLh7i}.} relies on the following intuitions: 
\begin{itemize}
	\item the states and the (probabilistic) transitions of the LTS behind the MDP are constructed based on the variables, actions and events in the \react domain model;
	\item the rewards in the MDP are associated with transitions that lead to the satisfaction of requirements.
\end{itemize}

\paragraph*{Dealing with the Markovian assumption} Building an MDP from an \react model requires the satisfaction of the Markovian assumption. In the context of this work, determining the satisfaction of requirements, with the exception of unconditional requirements, requires to keep track of history. To solve this issue, we extend the state space to store information that is relevant to determine the status of requirements in every state. This is done by associating every requirement with a state variable, whose value reflects the status of the requirement in the state\footnote{This technique is inspired by the state-based approach in~\cite{Bacchus96} to handle non-Markovian rewards, but is tailored to support requirements in \react.}. 
The value of those variables, called {\it requirements variables}, are updated whenever their corresponding requirement is activated, canceled, satisfied, etc.  

\textit{Requirements Variables $\mathcal{RV}$} are special variables whose domain represents all the possible statuses of their corresponding requirement. The statuses of requirements and their update after requirement activation, cancellation, satisfaction, etc are defined in the transitions part of Figures~\ref{fig:unconditionalRE} and \ref{fig:conditional_deadline}. On the other hand, the rewards part defines transitions that satisfy requirements and, consequently, entail a reward. 

For example, consider a conditional achieve requirement $CA$ of the form $\textbf{ReqID}\, m \,\textbf{achieve}\, S \,\textbf{if} \, A \,\textbf{unless} \, Z\,\textbf{reward} \, r$. This requirement is associated with a requirement variable $m$ whose domain includes the requirement's possible statutes $\{I,R\}$. 
The transitions part of Figure~\ref{fig:conditional_deadline} shows the evolution of $CA$ requirements when their activation, cancellation and required conditions occur. It is to be read as follows: when the status is $I$ and the activation condition $A$ is true, then the status is updated to $R$. Analogously, if the status is $R$ and the cancellation condition $Z$ or the required condition $S$ is true, then the status is updated to $I$. 
The updating of a state variable as just described enables the definition of a Markovian reward when requirements are satisfied. The rewards part of Figure~\ref{fig:conditional_deadline} shows transitions of $CA$ requirements that entail rewards. This figure should be read as follows: a requirement $m$ of type $CA$ induces a reward $r$ on a transition from a state $i$ to a state $j$ iff, in $i$, the required condition of $m$ does not hold and the status of $m$ is $R$; while $S$ holds in $j$.


\paragraph*{Dealing with the absence of exogenous events} \react models have explicit-event models whereas MDPs impose an {\it implicit-event action} model. To overcome this limitation, we exploit the technique proposed in~\cite{boutilier99} which enables computation of {\it implicit-event} action transition matrices from explicit-event models. The use of this technique assumes the following rules: 1) the action in which exogenous events are folded, always occurs before it and 2) events are {\it commutative}, i.e., their order of occurrence from an initial state produces the same final state. Under those assumptions, which are satisfied in our running example, the {\it implicit-event} transition matrix $Pr_a(s_i,s_j)$ of an action $a$ is computed in two steps: first, the transition matrix of $a$ (without exogenous events) and the transition matrix and occurrence vector of every event $e$ are computed separately; then, those elements are used to construct the implicit-event matrix of every action $a$. This process is illustrated in the following section using an example. Note that it is possible to integrate other (more complex) interleaving semantics into the framework if necessary by changing the technique used to compute the {\it implicit-event} transition matrix~\cite{boutilier99}. 


\subsection{MDP Construction Process}
An \react MDP $MDP_r=\langle \mathcal{S}, \mathcal{A}, \mathcal{T}, \mathcal{R}, \gamma \rangle$ is constructed from a model $\mathcal{D}_r=\langle \mathcal{SV}, \mathcal{AV}, \mathcal{ED}, \mathcal{RQ}, s_0 \rangle$ as follows. 

\textit{States $\mathcal{S}$: }
represent all possible configurations of the system and the environment. A state is a specific {\it configuration}, i.e., an assignment of every state variable in $\mathcal{SV}$ and requirement variable in $\mathcal{RV}$ a value from their domain. 

\begin{figure}[t]
	\centering
	\includegraphics[width=70mm, scale=0.7]{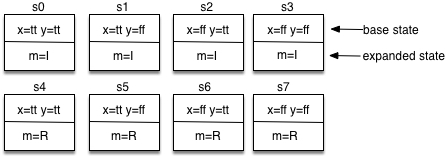}
	\caption{Constructed States}
	\label{fig:example-states}
\end{figure}

For example, consider a domain model $\mathcal{D}_r$ comprising of two boolean variables $x$ and $y$ and one requirement $m$ of type $CA$. The set of states $\mathcal{S}$ constructed from $\mathcal{D}_r$ comprises all possible configurations of its state and requirement variables. Thus, $\mathcal{S}$ includes the eight states in Figure~\ref{fig:example-states}.

\textit{Actions $\mathcal{A}$:} are all the actions appearing in $\mathcal{AV}$ of the domain model $\mathcal{D}_r$ extended with the empty action $noop$, which produces no effects and has no cost, i.e., $\mathcal{A}=\mathcal{AV}\cup\{noop\}$. 

\textit{The transition matrix $\mathcal{T}$:} is computed in two steps: first, the transition matrix of $a$ (without exogenous events) and the transition matrix and occurrence vector of every event $e$ are computed separately; then, those elements are used to construct the implicit-event matrix of every action $a$.

For example, consider that our domain model $\mathcal{D}_r$ includes one probabilistic action $a$, one deterministic action $b$ 
and one requirement $m$, which are defined as follows:
\begingroup\makeatletter\def\f@size{9}\check@mathfonts
\def\maketag@@@#1{\hbox{\m@th\large\normalfont#1}} 
\begin{align*}
&\textbf{Action } a \textbf{ if } !x \textbf{ effects }  \langle x \textbf{ prob } 0.8 \rangle \langle y \textbf{ prob } 0.2 \rangle \; \textbf{ cost } 10\\ 
&\textbf{Action } b \textbf{ if } x \textbf{ effects }  \langle !x \rangle \; \textbf{ cost } 5\\ 
&\textbf{ReqID } m \textbf{ achieve } x \textbf{ if }  !x \textbf{ reward } 100
\end{align*} \endgroup

\begin{figure}[t]
	\centering
	\includegraphics[width=\linewidth]{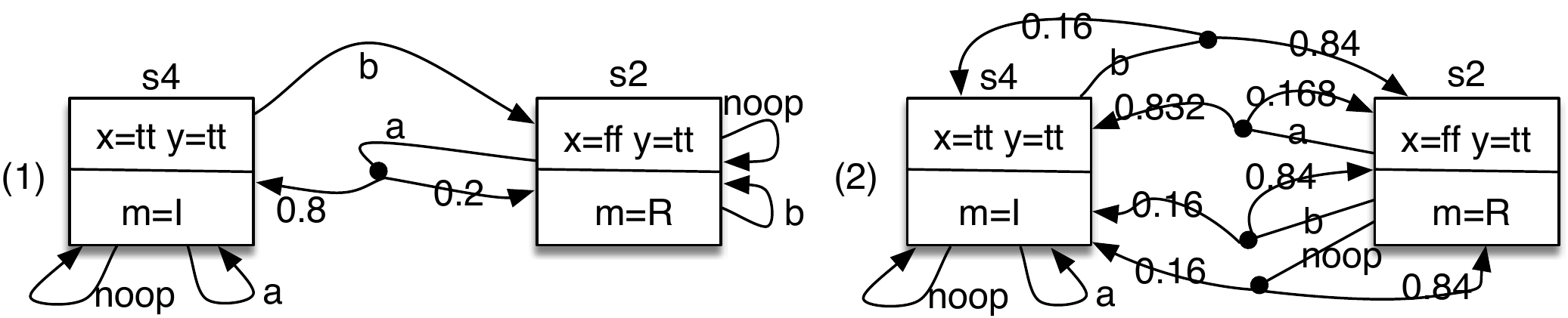}
	\caption{(1) action transitions, (2) implicit-event action transitions.}
	\label{fig:example-action-transition}
\end{figure}

\noindent Figure~\ref{fig:example-action-transition}(1) shows the transitions caused by the execution of actions in the states $s_2$ and $s_4$. For example, since the condition $!x$ is satisfied in $s_2$, the execution of $a$ in $s_2$ produces $x$ with a probability of $0.8$ and produces $y$ with a probability of $0.2$. Notice that after the execution of $a$, the base state of both $s_0$ and $s_4$ could be the result of executing action $a$ in $s_2$. However, since the execution of $a$ satisfies the requirement $m$, i.e., makes $x$ true, only the expanded state of $s_4$ satisfies the state transition model of the $CA$ requirement $m$ shown in Figure~\ref{fig:conditional_deadline} since $m=I$. Thus, the execution of $a$ in state $s_2$ leads to $s_0$ with a probability $0.8$, i.e., $Pr_a(s_2,s_4)=0.8$. The execution of $b$ and $noop$ do not change the state.



Events are similar to actions with the exception that they have occurrence probabilities, do not have a cost and do not advance time since they occur {\it concurrently} with actions. Let $e$ be an event defined similarly to $a$ as follows:
\begingroup\makeatletter\def\f@size{9}\check@mathfonts
\def\maketag@@@#1{\hbox{\m@th\large\normalfont#1}} 
\begin{align*}
&\textbf{Event } e \textbf{ if } !x \textbf{ occur prob } 0.2 \textbf{ effects }  \langle x \textbf{ prob } 0.8 \rangle \langle y \textbf{ prob } 0.2 \rangle 
\end{align*} \endgroup
\noindent In this case, the transition matrix of $e$ is similar to that of $a$, i.e., $Pr_e=Pr_a$. The occurrence vector $O_e$ of $e$ represents the probability of occurrence of $e$ in every state. Since the condition $\neg x$ is satisfied in the states $s_2$, $s_3$, $s_6$ and $s_7$, $O_e(s_2)=O_e(s_3)=O_e(s_6)=O_e(s_7)=0.2$. 
Figure~\ref{fig:example-action-transition}(2) shows the {\it implicit-event} transitions for the states $s_2$ and $s_4$: in $s_2$, event $e$ may occur with a probability of $0.2$, thus its effects are {\it factored in} action transitions as shown in~Figure~\ref{fig:example-action-transition}(2); in $s_4$, the condition of $e$ is not satisfied and, therefore, it does not affect the computed transitions for the actions $noop$ and $a$. Due to the interleaving semantics where exogenous events (may) occur after action execution, the transition caused by $b$ in $s_4$ is affected due to the possibility that $e$ occurs after $b$. 

\paragraph*{Construction of the reward matrix $\mathcal{R}$:} Transition rewards are affected by: (1) action costs and (2) satisfaction of requirements. In particular, a transition reward $R_a(s_i,s_j)$ is the sum of rewards obtained due to satisfaction of requirements on the transition from $s_i$ to $s_j$ minus the cost of $a$. For example, consider the transition from $s_2$ to $s_4$ caused by the execution of $a$ in $s_2$. On this transition, the  requirement $m$ is satisfied. Since the cost of executing $a$ is $10$, this transition will be associated with a reward of $100-10=90$, i.e., $R_a(s_2,s_4)=90$. 

\subsection{Requirements Transitions and Rewards}
This section explains the key intuitions behind the modeling of requirements in \react and their semantics.

\begin{figure}
	\centering
	\includegraphics[width=75mm, scale=0.75]{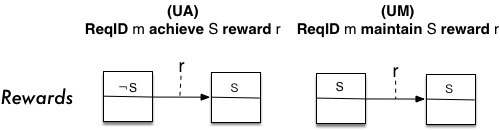}
	\caption{Unconditional Requirements States and Rewards}
	\label{fig:unconditionalRE}
\end{figure}

\paragraph*{Unconditional Achieve and Maintain Rewards} A maintain requirement defines a condition that should be kept satisfied. Therefore, a reward is given to the agent whenever this condition holds over two consecutive states, see, e.g., $\br{UM}$ in Figure~\ref{fig:unconditionalRE}. On the other hand, an achieve requirement defines a condition that should be reached. Therefore, the agent is rewarded when this condition becomes true, i.e., when it does not hold in a state but holds in the next, e.g., see $\br{UA}$. 

\begin{figure*}
	\centering
	\includegraphics[width=0.85\linewidth]{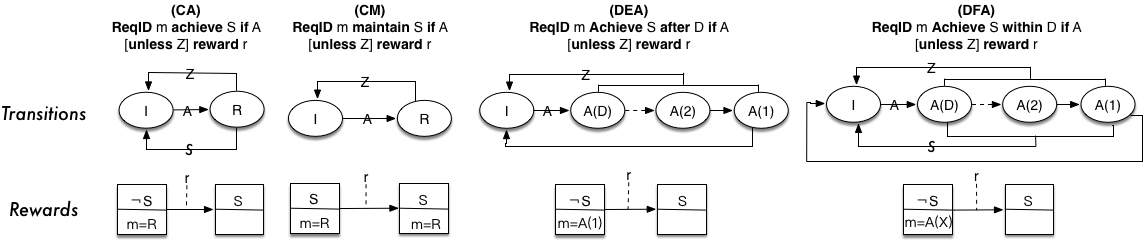}
	\caption{Conditional and Achieve Deadline Requirements States and Rewards}
	\label{fig:conditional_deadline}
\end{figure*}

\paragraph*{Conditional Requirements} The satisfaction of requirements is often necessary only after some condition $A$ becomes true, see for instance $\br{CA}$ and $\br{CM}$ in Figure~\ref{fig:conditional_deadline}. Those requirements are therefore modeled as state machines which are initially in an initial or inactive state $I$. When their activation condition $A$ occurs, a transition to a new state $R$ occurs. In a state $R$, the requirement is said to be {\it in force}, i.e., its satisfaction is required. While in $R$, the reward $r$ is obtained whenever the agent manages to comply with the required condition $S$. If the cancellation conditions $Z$ is detected while the requirement is in force, a transition to $I$ occurs, i.e., the requirement is canceled and has no longer to be fulfilled.

\paragraph*{Deadline Achieve Requirements} Requirements are sometimes associated with fixed deadlines. Fixed deadlines can represent either an exact time after which the agent should comply with the requirement, see for example $\br{DEA}$ in Figure~\ref{fig:conditional_deadline}; or a period of (discrete) time during which the agent may comply at any moment, see for example $\br{DFA}$. In both cases, deadlines are modeled similarly. For example, consider a requirement $m$ having a deadline $D$. After $m$'s activation, a transition to a state $A(D)$ occurs. At every subsequent time unit, a transition from a state $A(X)$ to a state $A(X-1)$ occurs (unless $X=1$). A transition entails the requirement's reward if the requirement is satisfied on this transition.

\comments{
	\begin{figure}
		\centering
		\includegraphics[width=\linewidth]{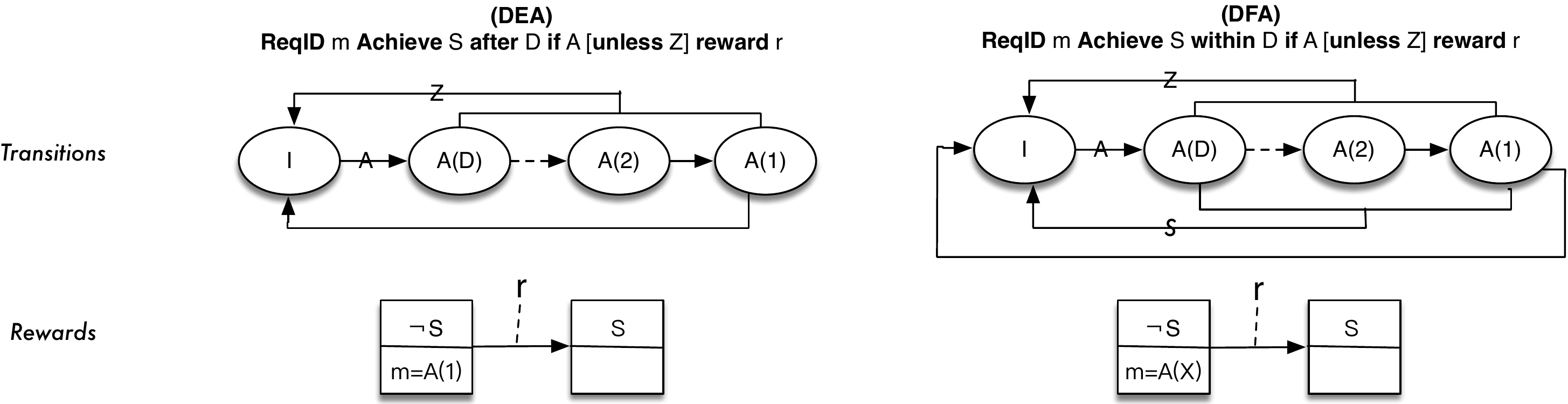}
		\caption{Deadline Achieve Requirements: State Machines and Associated Rewards}
		\label{fig:deadlineARE}
	\end{figure}
}

%% file: Evaluation.tex
\section{Evaluation}
\label{sec:evaluation}
In this section, we first present an empirical evaluation of the framework by comparing the use of an \react controller to control $RoboX$ in a simulated software environment of the restaurant $FoodX$ to a generic Monitor Analyze Plan Execute (MAPE) controller (Section~\ref{sec:empirical_evaluation}). The MAPE controller relies on a Planning Domain Description Language (PDDL) planner, similarly to state-of-the-art robotic systems such as ROS~\cite{rosplan}. 
Then, we present a qualitative comparison of \react with state-of-the-art probabilistic model-checkers, PAT~\cite{SunLDP09}, PRISM~\cite{Kwiatkowska2011} and STORM~\cite{Dehnert2017} which have been used in several other previous works~\cite{Camara2014a,Camara2015,Moreno2015,Camara2016} to solve adaptation problems (Section~\ref{sec:comparison_probabilistic_model_checkers}). Finally, we describe the current prototype tool implementation and conduct a performance evaluation (Section~\ref{sec:preliminaryeval}).

\subsection{Empirical Evaluation}
\label{sec:empirical_evaluation}
Figure~\ref{fig:simulation} depicts our simulation environment. It consists of a system state, an agent and an environment. The simulation runs in discrete time steps. At each step, the agent has to choose, based on the current system state, one action to execute from actions whose preconditions are satisfied in the state. On the other hand, some events are selected for execution, according to their occurrence probability, if their preconditions are true. After each time step, the state is updated by applying the effects of the chosen action and events in the current state. Effects of both actions and events are applied probabilistically according to the probabilities specified in their action/event descriptions, i.e. their execution can lead to different states. Experiments are run for one hundred thousands steps.

To select the agent's actions, two controllers were implemented: an \react controller and a generic MAPE controller. The design rational of the experiment aims at comparing a \react controller and generic MAPE controllers with respect to: types of supported requirements, enforcement model, response time, quality of decision-making and problem representation.

\paragraph*{Experiment Description} The experiment ran on a MacBook pro with a 2.2GHz Intel Core i7 and 16 GB of DDR3 RAM. At each time step, the agent queries the state (the (M)onitoring activity). The agent determines its action to execute by interacting with its controller. The controller, given the current state, determines the next action of the agent. 
The \react controller is implemented in Java and uses the computed \react strategies to determine the optimal action that the agent should take at each state. The MAPE controller is also implemented in Java. It consists of three components: 1) an analysis component that determines whether planning is needed, 2) PDDL4J, an open source Java library for Automated Planning based on PDDL~\cite{PDDL4J}, to compute plans and 3) a plan enforcer which returns one action at each step to the agent. Below is a comparison of the two controllers.

\paragraph*{Supported requirements} \react supports the types of requirements presented in Section~\ref{sec:modeling_requirements}. The MAPE controller, since it relies on a PDDL planer, can naturally encode unconditional and conditional achieve requirements, i.e. $\langle UA, CA\rangle$. The other types of requirements cannot be easily encoded in the form of PDDL planning problems. 

\paragraph*{Enforcement model} The \react controller enforces MDP strategies. It has a simple enforcement model: it consults the computed strategy and determines the optimal action that corresponds to the current state at each time step. The MAPE controller enforces requirements as follows: if the activation condition of a conditional requirement is true in the state then a planning problem (Pb) is formulated to satisfy the requirement's condition. When multiple requirements must be satisfied, then the goal state of the planning problem corresponds to the disjunction of their (satisfaction) conditions, i.e. one requirement should be satisfied. If a plan (P) is found by the PDDL planner, the plan enforcer module selects one action of P to return to the agent at each time step. A plan is pursued until its end, i.e. no re-planning is performed until the plan's last action is executed unless a plan execution fails. A plan fails if one of its actions cannot be executed because its preconditions are not satisfied in the current state. This situation may occur due to nondeterministic action effects or event occurrences\footnote{Another situation where re-planning would be required is cancellation of requirements. This situation is not considered in this experiment for simplicity.}.

\begin{figure*}[!htb]
	\minipage{0.32\textwidth}
		\includegraphics[scale=0.3]{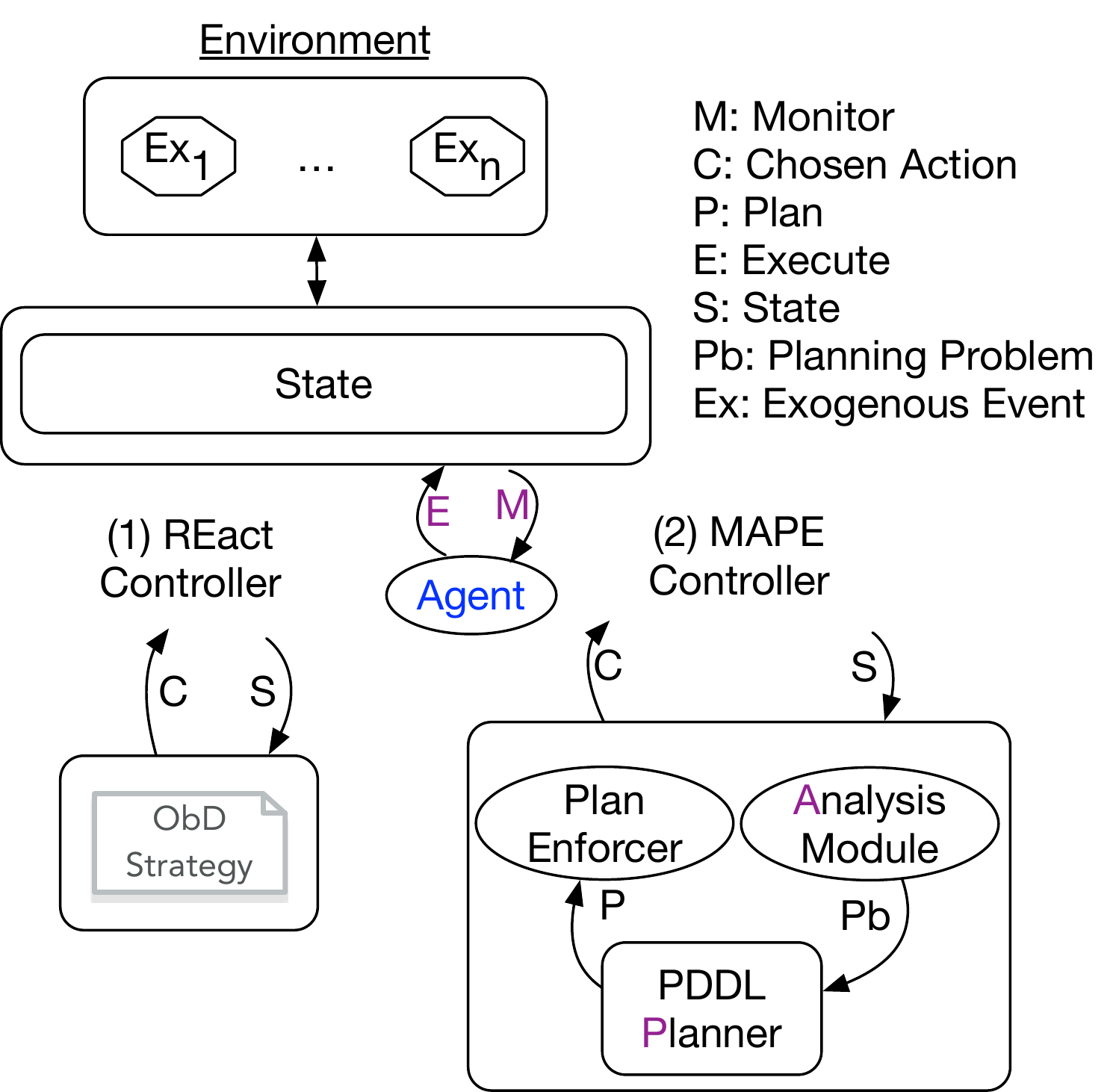}
		\caption{Experiment Description}\label{fig:simulation}
	\endminipage\hfill
	\minipage{0.32\textwidth}
		\begin{tikzpicture}[scale=0.6]
	\begin{axis}[ylabel=decision-making time per time step,xlabel=nb of planning goals]
	\addplot coordinates {
		(0,0)
		(4,261)
		(8,294)
		(12,288)
	};
	\addplot coordinates {
		(0,0)
		(4,284)
		(8,378)
		(12,370)
	};
	\addplot coordinates {
		(0,0)
		(4,348)
		(8,488)
		(12,463)
	};
	\addplot coordinates {
		(0,0)
		(4,409)
		(8,586)
		(12,556)
	};
	\addplot coordinates {
		(0,0)
		(4,514)
		(8,689)
		(12,639)
	};
	\addplot coordinates {
		(0,0)
		(4,593)
		(8,719)
		(12,724)
	};
	
	\addplot coordinates {
		(0,0)
		(4,0.2)
		(8,0.3)
		(12,0.25)
	};
	
	\addlegendentry{M-100\%}
	\addlegendentry{M-90\%}
	\addlegendentry{M-80\%}
	\addlegendentry{M-70\%}
	\addlegendentry{M-60\%}	
	\addlegendentry{M-50\%}
	\addlegendentry{A-all}
	\end{axis}
	\end{tikzpicture}

\caption{Decision-making time per time step}
\label{lab:fig_response}
	\endminipage\hfill
	\minipage{0.32\textwidth}%
	\begin{tikzpicture}[scale=0.6]
\begin{axis}[ylabel=nb of satisfied goals per time step,xlabel=nb of planning goals]
\addplot coordinates {
	(0,0)
	(4,0.2)
	(8,0.228)
	(12,0.24)
};
\addplot coordinates {
	(0,0)
	(4,0.152)
	(8,0.208)
	(12,0.216)
};
\addplot coordinates {
	(0,0)
	(4,0.14)
	(8,0.188)
	(12,0.192)
};
\addplot coordinates {
	(0,0)
	(4,0.128)
	(8,0.164)
	(12,0.168)
};
\addplot coordinates {
	(0,0)
	(4,0.116)
	(8,0.144)
	(12,0.124)
};

\addlegendentry{M-100\%}
\addlegendentry{M-90\%}
\addlegendentry{M-80\%}
\addlegendentry{M-70\%}
\addlegendentry{M-60\%}	
\addplot coordinates {
	(0,0)
	(4,0.2)
	(8,0.264)
	
};
\addplot coordinates {
	(0,0)
	(4,0.188)
	(8,0.244)

};
\addplot coordinates {
	(0,0)
	(4,0.176)
	(8,0.224)
	
};
\addplot coordinates {
	(0,0)
	(4,0.16)
	(8,0.204)
	
};
\addplot coordinates {
	(0,0)
	(4,0.144)
	(8,0.18)
	
};
%

\addlegendentry{A-100\%}
\addlegendentry{A-90\%}
\addlegendentry{A-80\%}
\addlegendentry{A-70\%}
\addlegendentry{A-60\%}	
\end{axis}
\end{tikzpicture}
\caption{Goal satisfaction per time step} \label{fig:rewards_mape_react}
	\endminipage
\end{figure*}

%
%
\paragraph*{Response time} Figure~\ref{lab:fig_response} shows the average time of decision making, i.e. the total decision-making time divided by the total number of steps of the experiment. Several domain descriptions differing in their total number of planning goals/requirements and action success rate\footnote{Action success means that the action produced its (expected) effect, i.e. the effect that is most likely to occur.} (100\%-50\%) are considered. The decision time of an \react controller is constant and almost instantaneous ($\sim$200ns) as it consists of a simple lookup in the policy (which is stored in the form of an array of Integers) of the optimal action that corresponds to the current state. On the other hand, the analysis and planning activities of the MAPE controller introduce a significant overhead when compared to the \react controller. The average decision-making time of MAPE controllers also increases with action non-determinism as re-planning is required more frequently due to more frequent plan failures.  

%
%
%
%
%
%
%
%

\paragraph*{Quality of decision-making} Figure~\ref{fig:rewards_mape_react} shows the number of satisfied goals per time step for MAPE and \react controllers. It demonstrates that the \react controllers consistently outperform MAPE controllers for the same domain problems. This is due, on one hand, to their ability to include probabilities of event occurrences into their computation of optimal strategies. For instance, imagine that $RoboX$ has to pass the order of a table to the kitchen but that it estimates that there is a high-likelihood that another table orders. In this case, the \react controller may delay moving to the kitchen to pass the order and wait until the other table orders first before passing the two orders to the kitchen together. MAPE controllers are incapable of incorporating such intelligence in their decision-making. Another reason explaining this result is that MAPE controllers, once a plan is computed, commit to it unless the plan fails to avoid getting stuck in re-planning without acting, which could happen if the frequency of change in the environment is high. This makes it impossible to guarantee the optimality of plans throughout their execution. On the other hand, \react strategies are guaranteed to always select the optimal action at each state.

\paragraph*{Representation} An important difference is the goal/requirement representation. In MAPE, planning goals have to be satisfiable using solely the actions that are available to the agent. For example, consider a goal to achieve that a table pays as many times as possible. The satisfaction of this goal requires interactions with the environment as described in Figure~\ref{fig:tablelts}. This requirement cannot therefore be expressed as a single planning goal but has to be decomposed into a set of planning goals, each of which has to be satisfiable by the actions available to the agent. On the other hand, thanks to the folding of events into actions, such requirements can be expressed directly in \reactt. Consequently, expression of requirements in \react can be much more succinct and enable system designers to focus of what should be satisfied rather than how they should be satisfied. In the running example, it was possible to represent four MAPE planning goals in the form of a single \react requirement.

\subsection{Qualitative Comparison of \react with State-of-the-Art Probabilistic Model-checkers}
\label{sec:comparison_probabilistic_model_checkers}
 State-of-the-art probabilistic model-checkers PAT~\cite{SunLDP09}, PRISM~\cite{Kwiatkowska2011} and STORM~\cite{Dehnert2017} support the description of various models using a variety of languages. In this work, we focus on MDP models because they support, as opposed to other models such as for example Discrete-Time Markov Chains, the synthesis of optimal strategies. With respect to the general-purpose languages proposed by probabilistic model-checkers, our model and language support exogenous events and various typical software requirements (Section~\ref{sec:modeling_requirements}), elements that cannot be modeled or expressed using the general-purpose MDP languages of PAT, PRISM or Storm. An extension of PRISM, namely PRISM-games, supports modeling of turn-based multi-player stochastic games. This enables the modeling of the environment as a separate player whose actions represent exogenous events. With respect to modeling of autonomous systems and their requirements, PRISM-games has two main limitations. First, similarly to PRISM, rewards have to be Markovain which means that there is no way to encode typical software requirements~\cite{Lamsweerde2009} such as those presented in Section~\ref{sec:modeling_requirements} using the provided (Markovian) reward structures. Furthermore, modeling of interactions between the agent and its environment in \react where multiple events occur with each action of the decision maker is more realistic and natural than, in turn-based PRISM-games, where the environment may only be represented in the form of a separate player who selects at most one event to execute after each action of the decision maker.

\comments{
\begin{figure}[t]
	\centering
	\includegraphics[width=0.8\linewidth]{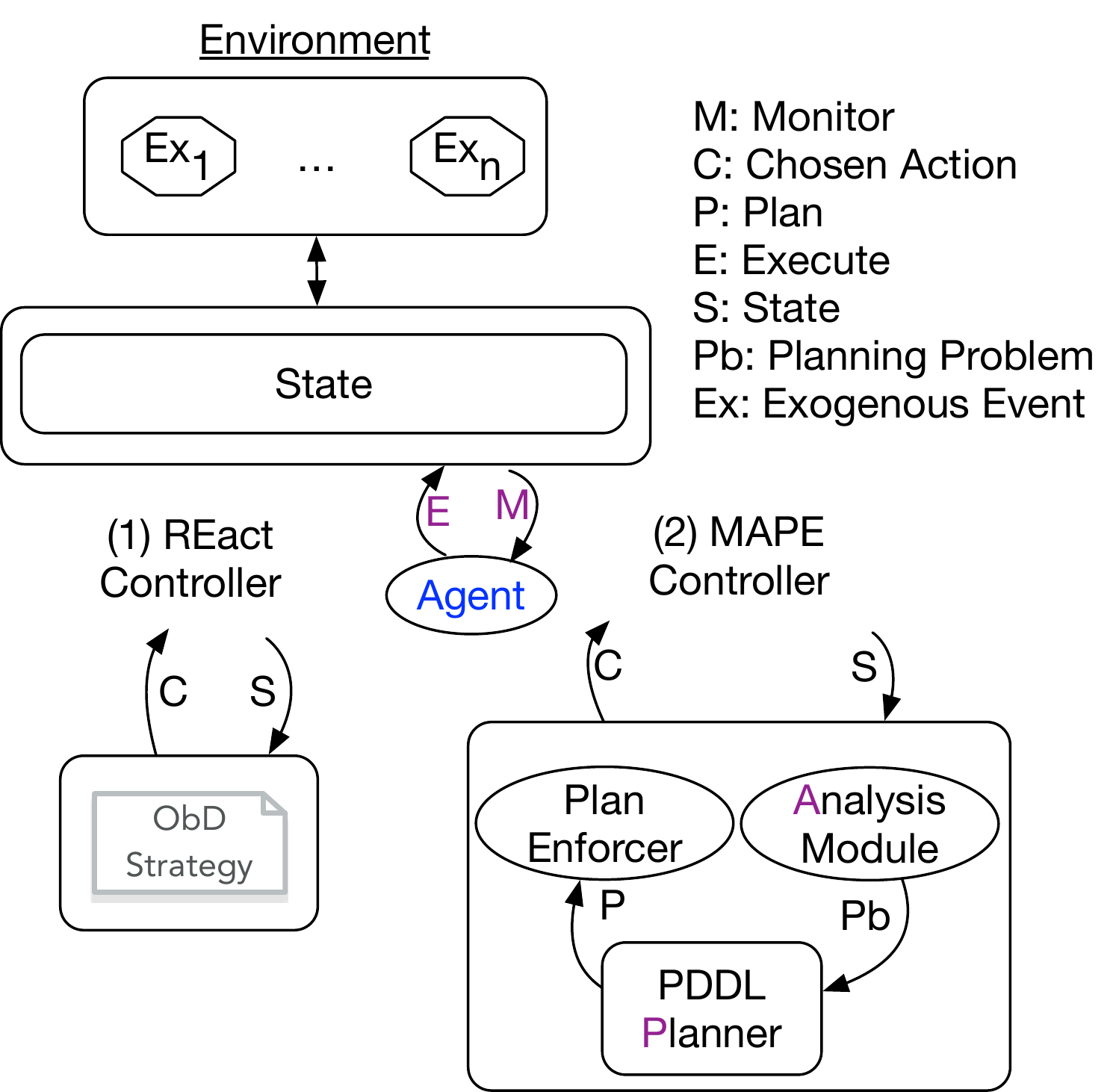}
	\caption{Simulation Description}
	\label{fig:simulation}
\end{figure}

\subsection{The benefits of \react}
\label{sec:action_costs}

As discussed in Section~\ref{sec:running_example}, many of the difficulties of adaptation in autonomous systems stem from uncertainties in the system, its environment and conflicts that arise when multiple requirements have to be satisfied at the same time. This section focuses on showing how \react deals with those challenges through use case examples.

\paragraph*{Dealing with Uncertainty}

\react allows the modeling of two sources of uncertainty: (1) action/event effects, and (2) occurrence of exogenous events. 

\textit{Effects} For example, imagine that $RoboX$ has to move to the office, which can be accessed in two ways: 1) directly from $table_2$ through a small door but that due to the size of the door, $RoboX$ succeeds to pass through the door only \textit{Pb}\% of the time, 2) indirectly by moving to the dining room, then to the hallway and finally into the office. This can be modeled by adding an action {\it move\_from\_table\_2} and a requirement $req\_1$ to the domain description:
\begin{align*}
	&\textbf{ReqID } req_1 \textbf{ achieve } location=inOffice\\ 
		&\textbf{Action } move\_from\_table_2 \textbf{ if } location=atTable_2\\
	& \hspace{1cm}  \textbf{ effects } \langle location=inOffice  \textbf{ prob } Pb\rangle 
\end{align*}

\noindent 
The computation of an optimal strategy determines that for every $Pb < .33$, $RoboX$'s optimal course of action is to take the longer route since it would be expected that fewer steps would be needed to move into the office. For every $Pb>=.33$, moving directly into the office from $table_2$ is the optimal action.  

\textit{Exogenous Events} In the running example, several exogenous events may occur: a customer may arrive, they may request to place an order, they may order, pay or leave. Those exogenous events may have preconditions. For example, it is possible that when new customers arrive, they occupy the first available table. For example, customers sit at $table_1$ if it is available, i.e., $table_1=empty$. Otherwise, they move to the next available table. This may be expressed as follows:

\begin{align*}
	&\textbf{Event } customer\_arrives\\
	& \hspace{0.5cm} \textbf{ if } table_1=empty \textbf{ occur prob } 0.1 \textbf{ effects } \langle table_1=occupied  \rangle \\
	& \hspace{0.5cm} \textbf{ if } !table_1=empty \,\&\, table2=empty \textbf{ occur prob } 0.1 \\
	& \hspace{0.7cm} \textbf{ effects } \langle table_2=occupied  \rangle \\
	& \hspace{0.5cm}...\\
	&\textbf{Event } customer\_orders_i \textbf{ if } table_i=requested \,\&\, location=\\
	& \hspace{0.5cm}  atTable_i \textbf{ occur prob } 0.8 \textbf{ effects } \langle table_i=received \textbf{ prob } 0.8 \rangle\\
	&\textbf{Event } customer\_pays_i \textbf{ if } table_i=delivered \\
	& \hspace{0.5cm} \textbf{ occur prob } 0.2\textbf{ effects } \langle table_i=paid  \rangle \\
	&\textbf{Event } order\_ready_i \textbf{ if } table_i=in\_preparation \\
	& \hspace{0.5cm} \textbf{ occur prob } 0.2\textbf{ effects } \langle table_i=ready  \rangle 
\end{align*}


Now, consider the following unconditional achieve requirement:
\begin{align*}
&\textbf{ReqID } req_2 \textbf{ achieve } table_1=paid 
\end{align*}

\noindent This requirement specifies that the ``payment'' state should be achieved {\it as many times as possible}. Figure~\ref{fig:scenario} shows parts of the policy computed for this domain description: it shows that $RoboX$ should move to $table_1$ and wait until customers arrive (see state $s_2$). After a customer arrives, it should ask them to order. Note that due to the chosen interleaving semantics, it could occur that the customer places an order simultaneously with the request (see the transition from $s_3$ to $s_5$). Once order is received, $RoboX$ must move to the kitchen, collect the order and then deliver the order back to the customer. 
After the customer leaves, the strategy determines that $RoboX$ should clean the table and wait for new customers. 

This strategy arguably defines a complete behavioral model for $RoboX$. The computation of this {\it optimal} strategy only requires a description of the environment, $RoboX$'s capabilities and the definition of requirements that capture the high-level goals that should be pursued. This example thus shows that suitable modeling and choice of requirements enables {\it succinct model descriptions}, based on which complex but optimal behavioral models can be derived. This should be contrasted with an approach where behavior of the agent has to be fully specified. The former approach is obviously more powerful since it: 1) requires much less effort, 2) makes explicit the {\it rational} {\it justifying} the agent's behavior and 3) provides optimality guarantees. 

\begin{figure}[t]
	\centering
	\includegraphics[width=0.9\linewidth]{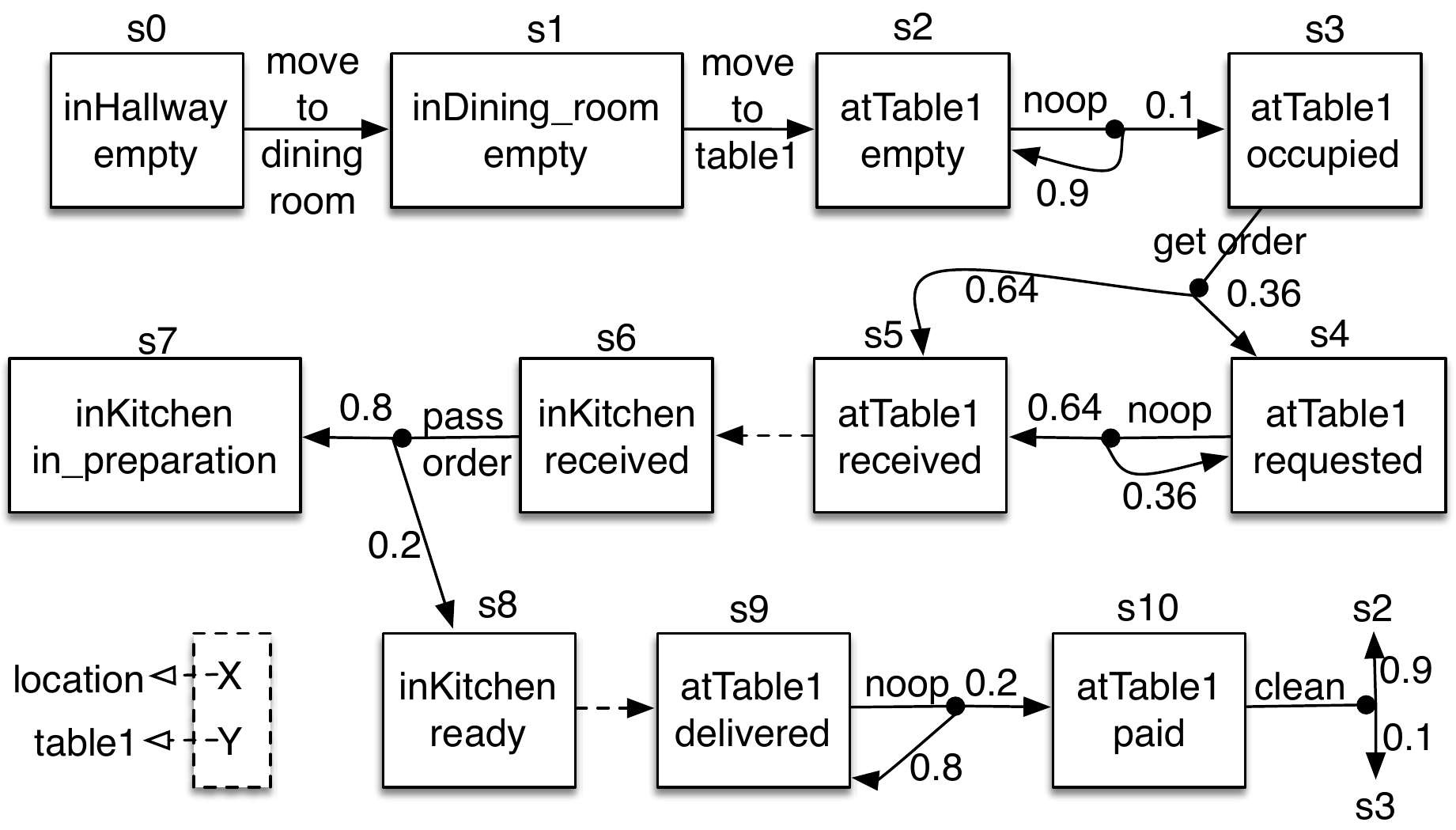}
	\caption{Computed Strategy Example (Selected States)}
	\label{fig:scenario}
\end{figure}

\paragraph*{Requirements Prioritisation}
When domain models include multiple requirement, the computed strategy prioritizes or schedules them according to the total reward that their satisfaction is expected to yield. For example, we extend the domain model with:
\begin{align*}
	&\textbf{ReqID } req_3 \textbf{ achieve } table_2=paid 
\end{align*}

\noindent Now, $RoboX$ has to serve $table_1$ and $table_2$. Consequently, many situations arise where $RoboX$ has to choose whether to pursue $req_2$ or $req_3$. Figure~\ref{fig:requirementscheduling} shows one of those situations: if $RoboX$ had gotten $table_1$'s order and customers at $table_2$ have not ordered yet, should it pass the order of $table_1$ to the kitchen or wait and get $table_2$'s order first? The optimal action in this case is to move to $table_2$ to get its order first before passing the two orders to the kitchen. Notice that this is indeed the optimal course of action according to the model description which specifies that $table_2$ is very likely to place an order.

Notice that {\it prioritisation} situations, similar to the one just discussed, do not have to be identified and resolved separately, as would be required when the behavior of the system is explicitly specified. A second benefit of our requirements-driven approach is that conflict resolution is {\it optimal} according to the domain model.

\begin{figure}[t]
	\centering
	\includegraphics[width=\linewidth]{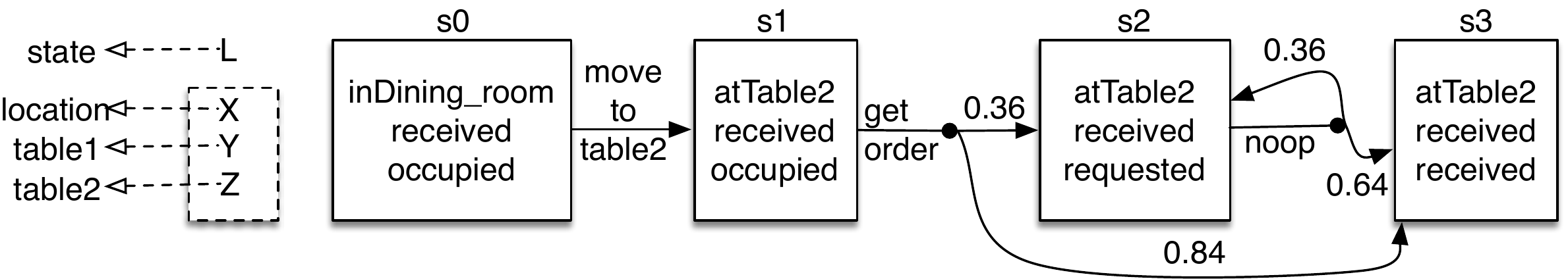}
	\caption{Requirement Prioritisation/Scheduling}
	\label{fig:requirementscheduling}
\end{figure}

\paragraph*{Requirements Tradeoffs}
\label{sec:requirements_tradeoffs}
When requirements have deadlines, i.e. must be complied with within a finite number of time units, scheduling/prioritisation may not be possible. Notice that without deadlines, some requirements could remain unsatisfied indefinitely. We currently identify those situations by inspecting the generated strategies. The automation of the detection of those cases is left for future work. 
Now, consider the following requirements:
\begin{align*}
	&\textbf{ReqID } req_4 \textbf{ achieve } table_1=delivered  \textbf{ within } 8 \\
	&\hspace{0.5cm} \textbf{ if } order_1=ready \textbf{ reward } 1000\\		
&\textbf{ReqID } req_5 \textbf{ achieve } table_2=empty \textbf{ within } 2 \\
&\hspace{0.5cm} \textbf{ if } order_2=paid \textbf{ reward } 500
\end{align*}

\begin{figure}[t]
	\centering
	\includegraphics[width=\linewidth]{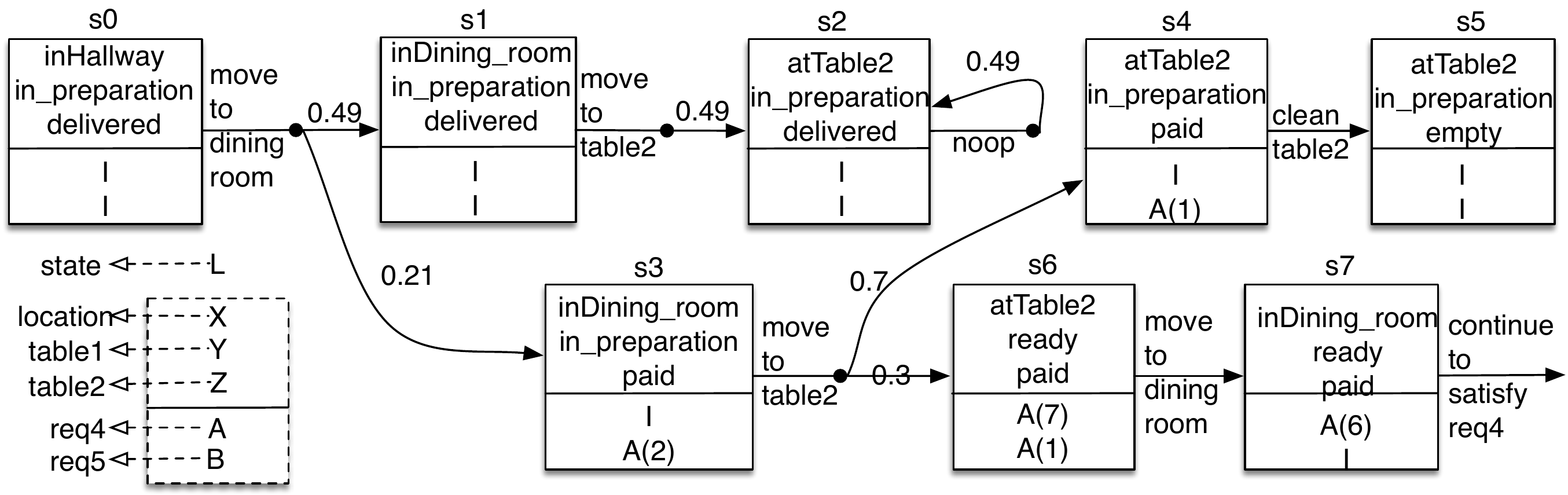}
	\caption{Requirements Trade-offs}
	\label{fig:tradeoffs}
\end{figure}
\noindent Figure~\ref{fig:tradeoffs} shows the optimal actions for seven of the possible states. They show the following features of computed strategies:

\textit{Proactive behavior:} In $s_0$, $RoboX$ does not have requirements that are {\it active} or {\it in force}. It however anticipates the activation of $req_5$ which have to be satisfied within 2 time steps. Therefore, the optimal action is to move to $table_2$ (see the states $s_0$-$s_2$). 

\textit{Disciplined Trade-offs:} Consider the situation in state $s_6$: $RoboX$ is at $table_2$, it has to clean $table_2$ (requirement $req_5$ is active) but at the same time $table_1$'s food is ready (requirement $req_4$ is active). In this case, the optimal action is to abandon the satisfaction of $req_5$ since it conflicts with the satisfaction of $req_4$, which reward is higher. Note that the collection and delivery of food starting from $table_2$ takes 8 time units, the exact time specified in the deadline of $req_4$. Hence, $RoboX$ cannot satisfy both requirements.

}
\subsection{Preliminary Experimental Evaluation}
\label{sec:preliminaryeval}
We have implemented the \react framework as a Java-based prototype which uses EMFText~\cite{Heidenreich2013}, the MDPToolBox package~\cite{MDPToolbox2017} and Graphviz~\cite{Ellson2002}.
\comments{
\begin{itemize}
	\item EMFText~\cite{Heidenreich2013} is a model-based technique for the design and implementation of domain specific languages (DSLs) based on the Eclipse Modeling Framework (EMF) and its underlying modeling language Ecore. EMFText is used to define the meta-model and textual representation ({\it concrete syntax}) of \react presented in Section~\ref{sec:model_language};
	\item MDPToolBox package~\cite{MDPToolbox2017} provides classes and functions for the resolution of discrete-time MDPs in MATLAB. 
	\item Graphviz is used for the visualization of the constructed MDP models and their solutions, i.e., the computed strategies, enabling the manual inspection and verification of adaptation strategies. The identification of properties of sound adaptation strategies and the automation of their verification represents future work.
\end{itemize}
}
There are at least two main use cases of the framework:
	\paragraph*{At design-time} the textual editor generated by EMFText can be used to define \react models. The corresponding MDP models and optimal strategies can be then visualized and inspected by a system designer and/or used to synthesize optimal controllers for the target autonomous systems;
	\paragraph*{At runtime} the \react Java API can be used to create instances of the \react model and the computation of optimal strategies at runtime. At runtime, strategies should be recomputed after change in either 1) requirements or 2) domain descriptions. The former generally denotes a change in system objectives or their priorities. On the other hand, the latter is needed if new information (possibly based on interactions with the environment) shows that model parameters need to be revised. There are some limitations to this use scenario which are discussed in Section~\ref{sec:limitations}.

\begin{figure}
	\centering
	\begin{minipage}{.5\linewidth}
		\centering
	\includegraphics[width=\linewidth]{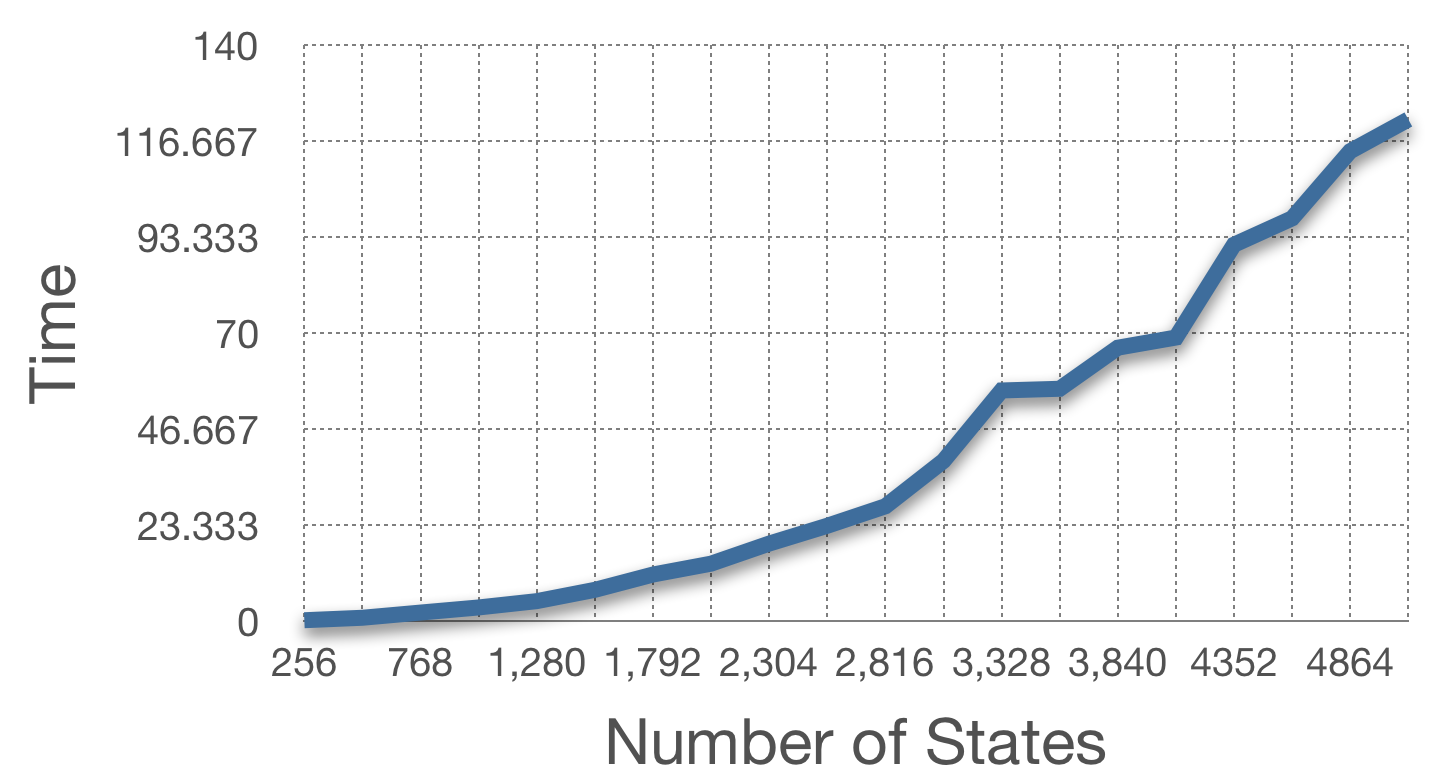}
		\caption{MDP Construction}
		\label{fig:mdp_construction_time}
	\end{minipage}%
	\begin{minipage}{.5\linewidth}
		\centering
	\includegraphics[width=\linewidth]{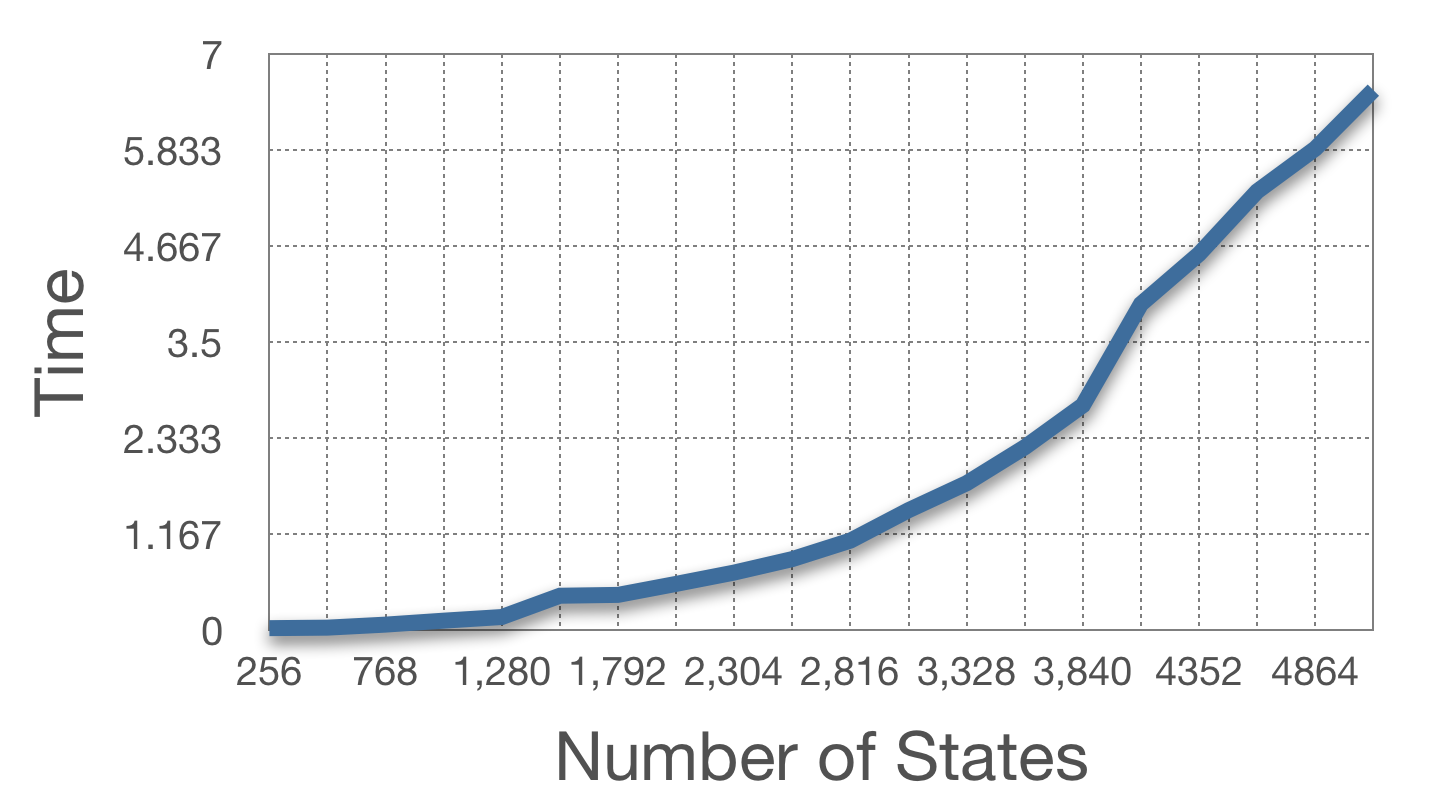}
		\caption{MDP Solution}
	\label{fig:mdp_solution_time}
	\end{minipage}
\end{figure}

Figures~\ref{fig:mdp_construction_time} and \ref{fig:mdp_solution_time} show the MDP construction and solving time for different state space sizes, respectively. It is clear from the figures that the current implementation suffers from the state explosion problem. However, the support of thousands of states is typically sufficient for a large number of problems. Furthermore, solving an MDP is a one-time effort, i.e. once an MDP is solved (given a set of requirements), the computed strategies can be used until either requirements or the domain model change.
The improvement of the performance of our current prototype represents future work. 

%% file: Limitations.tex
\section{Limitations \& Issues}
\label{sec:limitations}
This section discusses the current limitations and issues related to using \react and means to address them.
\paragraph*{Setting of Model parameters} determining probabilities of actions and events can be challenging. We envisage that they shall be computed by adapting existing techniques that enable computation and learning of model parameters at runtime. For example, we could use~\cite{Epifani2009} where Bayesian techniques are used to re-estimate probabilities in formal models such as Markov chains based on real data observed at runtime; or~\cite{Calinescu2014} which proposes an on-line learning method that infers and dynamically adjusts probabilities of Markov models from observations of system behaviour. Alternatively, reinforcement learning techniques~\cite{Tesauro2007} could be used. 
	
\paragraph*{Identification of requirements} strategies are computed according to requirements. It is thus crucial that they be correctly identified. \react supports traditional goal modeling techniques~\cite{Lapouchnian2005,VanLamsweerde2009,Yu2011a}. Those techniques have been proven reliable over the years in ensuring correct elicitation, refinement, analysis and verification of requirements~\cite{Lapouchnian2005,VanLamsweerde2009,Yu2011a}. 
	
	\paragraph*{Suitability of the Application Domain} it is necessary to identify system conditions under which the framework may be used. Towards answering this question, we first define a predictable (unpredictable) system as one where probabilities of occurrence and effects of events/actions do not (do) change with time. Similarly, we define a dynamic (erratic) system as one where the rate of {\it relevant}\footnote{A change is relevant if it renders computed strategies obsolete.} change in those probabilities is within the order of hours or days (minutes or seconds). Our current prototype implementation computes strategies within minutes. Consequently, we conjecture that it supports the runtime synthesis of controllers in predicable and unpredictable dynamic systems. Erratic systems are not supported. A more precise definition of those limitations represents future work.



%% file: RelatedWork.tex
\section{Related Work}
\label{sec:related_work}

\begin{table}[t]
	\caption{Comparison of \react with Related Frameworks}
	\label{tab:comparison_table}
	\begin{tabular}{|c|c|c|c|c|c|c|c|c|}
		\hline 
		{\bf Comp.}    		 & Sub-criteria	&$F_1$&$Q_1$&$K_1$&$R_1$&$K_2$&$A_1$&$R$\\
		\hline	 
		\multirow{3}{*}{Model}       & Requirements &\yes &\yes &\yes &\yes &\yes &\yes  &\yes  \\
									 & Capabilities &\yes &\ps  &\yes &\ps  &\ps  &\ps  &\yes \\
								 	 & Events       &\no  &\ps  &\no  &\no  &\ps  &\no  &\yes \\
		
		\hline 	 
		\multirow{2}{*}{Uncert.}     & Occurrence   &\no  &\ps  &\no  &\no  &\ps  &\no  &\yes \\
								     & Effects      &\no  &\ps  &\no  &\yes &\ps  &\no  &\yes \\
		\hline 
		\multirow{3}{*}{Adapt  }     & Explicit     &\yes &\na  &\yes &\yes &\na  &\yes  &\na  \\
									 & Configuration&\na  &\yes &\na  &\na  &\yes &\na &\na  \\									
									 & Behavior     &\na  &\na  &\na  &\na  &\na  &\na  &\yes \\
		
		\hline 
		\multicolumn{9}{|l|}{\yes: supported \hspace{3.3cm} \no: not supported \hspace{1cm}}\\
		\multicolumn{9}{|l|}{\ps: partially supported (implicit) \hspace{1cm} \na: not applicable}\\
		\hline
		\multicolumn{9}{|l|}{$F_1$: FLAGS~\cite{Baresi2010}\hspace{0.5 cm}$Q_1$: QoSMOS~\cite{Calinescu2010}\hspace{0.5 cm}$K_1$: KAOS~\cite{VanLamsweerde2009,Letier2002,Cailliau2017}}\\
		\multicolumn{9}{|l|}{$R_1$: Rainbow~\cite{Garlan2004,Cheng2012}\hspace{2.3cm}$K_2$: KAMI~\cite{Filieri2012m}}\\
		\multicolumn{9}{|l|}{$A_1$: ActivFORMS~\cite{Iftikhar2014}\hspace{2.35cm}$R$: \react}\\
		
		\hline
	\end{tabular}
\end{table}

Table~\ref{tab:comparison_table} compares the features of \react with some notable requirements-driven adaptation frameworks according to the criteria in Sec.~\ref{sec:running_example}, divided along the following dimensions.

\textit{Modeling} compares the frameworks with respect to their support of the explicit modeling and representation of requirements, capabilities and events. Those feature are desirable as they simplify system design, its maintenance and modularity. 

\textit{Uncertainty} compares the support of uncertainty in exogenous event occurrences and effects. 

\textit{Adaptation} compares the type of adaptation strategies, which can be explicitly defined, configuration selection or behavior optimization. Configuration selection is a reactive approach where, after requirements are violated, the alternative system configurations are compared and the best one is selected. 
Behavior optimization is a proactive approach which takes into account not only the current conditions, but how they are estimated to evolve~\cite{Camara2014a}. 
Only behavior-based optimization supports the two requirements of (1) proactive and long-term behavior optimization, and (2) fast and optimal response to change. Note that adaptation based on explicitly defined strategies is fast but provides no optimality guarantees.

Table~\ref{tab:comparison_table} shows that adaptations in many current frameworks are either explicitly defined~\cite{Letier2002,Garlan2004,Lamsweerde2009,Baresi2010,Cheng2012,Iftikhar2014,Cailliau2017} or determined based on a comparison of possible system configurations~\cite{Calinescu2010,Filieri2012m}, without taking into account future evolutions of the system. It also shows that explicit-event and action models are rarely considered. For example, QosMOS and KAMI consider Markov chains. This is why these frameworks have an {\it implicit} models of actions and events in Table~\ref{tab:comparison_table}. 
Similarly, ActivForms rely on Timed Automata and the {\it Execute} activity is explicitly defined. Therefore, ActiveForms has explicit adaptation strategies and uncertainty is not handled. 
In \cite{Moreno2015}, MDP is used to identify optimal adaptations at runtime, taking into account the delay or latency required to bring about the effects of adaptation tactics . In~\cite{Camara2014a,Camara2016}, latency-aware adaptation is studied using stochastic multi-player games (SMGs). 
In \cite{Camara2015}, SMGs are used to generate optimal adaptation plans for architecture-based self-adaptation. These works exploit PRISM and PRISM-games to solve adaptation problems. So, they have the limitations discussed in Section~\ref{sec:comparison_probabilistic_model_checkers}.



Several other works~\cite{Esfahani2011,Elkhodary2010,Bencomo2013} studied the optimization of system configurations. In contrast, this paper focuses on behavior optimization. Several recent proposals explored the application of concepts from control theory~\cite{Filieri2014,Filieri2015,Shevtsov2016,Filieri2017} to perform system adaptation. One main difference with respect to these works is that their focus is on the optimization of quantifiable and measurable non-functional goals, such as response time, as opposed to behavior optimization based on functional requirements, the primary focus here.

%% file: Conclusion.tex
\section{Conclusion}
\label{sec:conclusion}
This paper introduces the \react framework for the model-based-require\-ments-driven synthesis of {\it reflex} controllers for autonomous systems. 
The framework introduces a model and a language to describe autonomous systems, their environment and requirements. The semantics of the model is defined in the form of an MDP, which can be solved producing optimal adaptation strategies (reflex controllers) for autonomous systems. In comparison with the general-purpose languages proposed by probabilistic model-checkers, \react solves two main limitations, namely the Markovian assumption and the implicit-event model. This enables the support of a comprehensive set of software requirements and permits the accurate modeling of the environment in which autonomous systems operate. 



Future work consists of 
extending the framework to support 
 online learning (reinforcement learning)~\cite{Tesauro2007}. The study of formal adaptation guarantees and assurances~\cite{Calinescu2017,Weyns2017a,Iftikhar2017a,Filieri2016}, and optimizing the performance of our framework~\cite{Pandey2016,Moreno2016} 
 are other future planned extensions. 

%% file: AppendixB.tex
\appendix

This appendix presents formally the mapping of \react model descriptions into an MDP. 

A model description is a tuple $\mathcal{D}_r=\langle \mathcal{SV}, \mathcal{AV}, \mathcal{ED}, \mathcal{RQ}, s_0 \rangle$, an $MDP_{r}=\langle \mathcal{S}, \mathcal{A}, \mathcal{T}, \mathcal{R}, \gamma \rangle$ is the MDP built on the basis of $\mathcal{D}_r$ if its various elements are constructed as described below. 

\paragraph*{\textit{State Atoms}}
	Let $\mathcal{SV}=\{x_1, ..., x_n \}$ be the set of state variables of $\mathcal{D}_r$ and $\{dom(x_1),...,dom(x_n)\}$ be their corresponding domains. 
	An assignment of a value $v_i\in dom(x_i)$ to a state variable $x_i$ is called a {\it state atom} over $x_i$. The set of state atoms $\mathcal{SA}=\{x_i$=$v_j \,|\, x_i\in \mathcal{SV} \textsl{ and } v_i\in dom(x_i)\}$ is called the set of {\it state atoms} of $\mathcal{D}_r$.

\paragraph*{\textit{Requirement Variables and Requirement Atoms}} Let $\mathcal{RQ}$ be the requirements of $\mathcal{D}_r$. 
	Let $r\in \mathcal{RQ}$ be a requirement and $name(r)$, $type(r)$ and $states(r)$ be functions returning the requirement's identifier, type and its possible states respectively. For example, let $r=\textbf{ReqID } m \textbf{ achieve } S \textbf{ if } A$ $ \textbf{ unless } Z \textbf{ reward } r$. In this case, $name(r)=m$, $type(r)=CA$ and $states(r)=\{I, R\}$. The set of requirements variables of $\mathcal{D}_r$ is the set 
	$\mathcal{RV}=\{r_1,...,r_m \}$ such that every $r_i$ is the name of a different requirement in $\mathcal{RQ}$. The domain of every $r_i$ is its set of possible states, i.e., $dom(r_i)=states(r_i)$.
	 The set $\mathcal{RA}=\{m=s \,|\, m\in\mathcal{RV} \textsl{ and } s\in dom(m)\}$ is called the set of {\it requirement atoms} of $\mathcal{D}_r$.

\begin{definition}[States $\mathcal{S}$] Let $\mathcal{V}$ be the set $\mathcal{SV}\cup \mathcal{RV}$. In this case, the set of states generated from $\mathcal{V}$ is the set $\mathcal{S} = \{\bigcup_{i=1}^{|\mathcal{V}|} \{x_i=v_i\} \,|\, x_i\in \mathcal{V} \, and \, v_i\in dom(x_i)\}$. 
\end{definition}

\noindent Intuitively, the previous definition means that every state $s\in \mathcal{S}$ is a set of atoms such that a value is assigned to every variable $x\in \mathcal{V}$. A state $s$ includes both state and requirements atoms. We distinguish between them as follows: state atoms of a state $s$ are referred to as the {\it base state} of $s$, denoted $\overline{s}$, and requirement atoms are referred to as the {\it expanded state} of $s$, denoted $\dot{s}$. More formally, let $s$ be a state, $\overline{s}=\{at \,|\, at\in s, \, at\in \mathcal{SA}\}$, whereas $\dot{s}=\{at \,|\, at\in s, \, at\in \mathcal{RA}\}$. This distinction is needed as state and requirements are updated differently: state atoms are directly updated by occurrence of actions and events; on the other hand, requirements atoms are indirectly updated if their status need be updated as a result of change in state atoms.

\paragraph*{\textit{Action Representation}} Let an action expression $ad\in \mathcal{AV}$ be a tuple $ad=\langle a,cost,\langle pre_1, \langle EF^1_1, p^1_1 \rangle,$ $..., \langle EF^1_m,$ $p^1_m \rangle \rangle,..., \langle pre_n,$ $ \langle EF^n_1,$ $ p^n_1 \rangle,..., \langle EF^n_l, p^n_l \rangle \rangle\rangle$ where $a$ is the action name, $cost$ its cost, $pre_i$ is one of its preconditions, every $\langle EF^i_x, p^i_x \rangle$ is one effect $x$ of the execution of $a$ when the precondition $pre_i$ holds and $p^i_x$ is the probability of producing the effect $x$. 

\begin{definition}[Actions] 
	The set of actions $\mathcal{A}$ is the set of action names in $\mathcal{AV}$ and the {\it noop} action, i.e., $\mathcal{A}$ is $\{a\,|\, \langle a,cost,\langle pre_1, \langle EF^1_1, p^1_1 \rangle,$ $..., \langle EF^1_m,$ $p^1_m \rangle \rangle,..., \langle pre_n,$ $ \langle EF^n_1,$ $ p^n_1 \rangle,..., \langle EF^n_l, p^n_l \rangle \rangle\rangle \in \mathcal{AV} \} \cup \{noop\}$ where $noop$ is an action which execution has no cost and produces no effects.
\end{definition}

\paragraph*{\textit{Formula Satisfaction}} Let $f$ be a formula of the form $6$ and $Y$ a set of atoms. The satisfaction of a formula $f$ in $Y$, denoted $Y\models f$, is defined in the usual way as follows:
	\begin{eqnarray*}
		&Y \models at              &  \textbf{ iff }  at\in Y \textbf{ otherwise } Y\not\models at\\ 
		&Y \models !f              &  \textbf{ iff }  Y\not\models f \\ 
		&Y \models f_1 \,\&\, f_2 &  \textbf{ iff } Y\models f_1 \text{ and } Y\models f_2\\
		&Y \models f_1 \,||\, f_2 &  \textbf{ iff } Y\models f_1 \text{ or } Y\models f_2
	\end{eqnarray*}

\begin{figure*}[t]
	\centering
	\includegraphics[width=\linewidth]{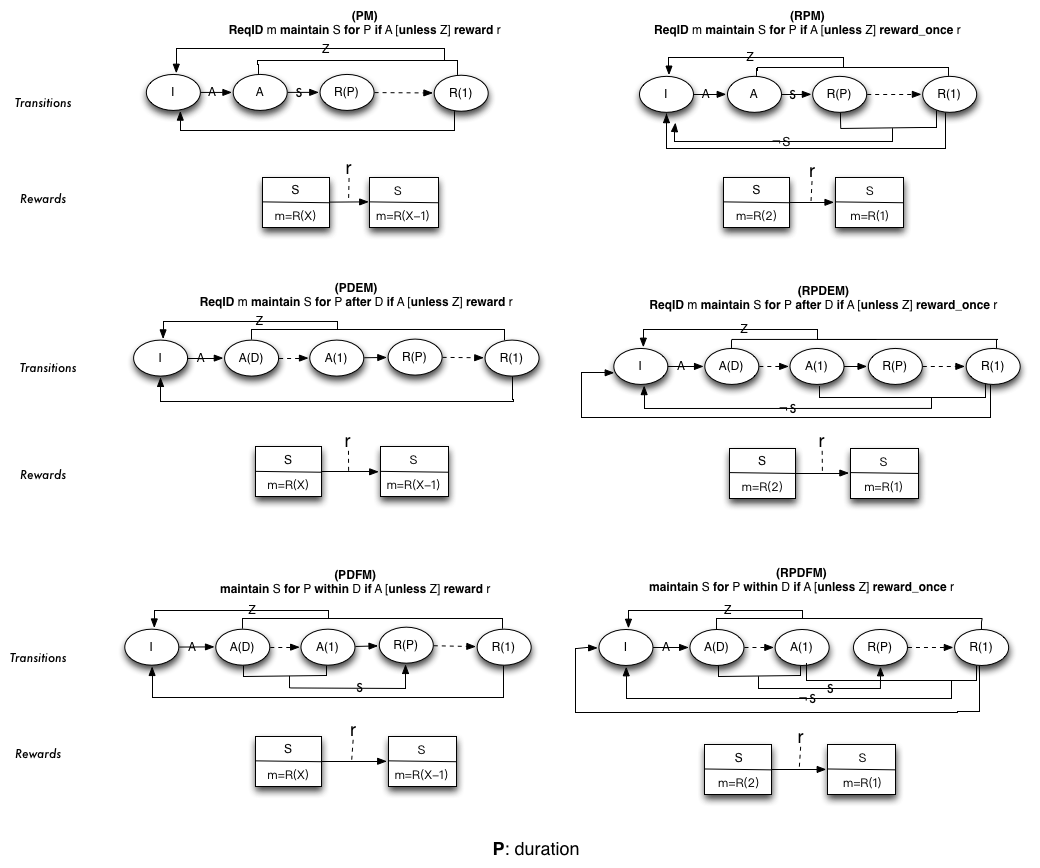}
	\caption{Duration Requirements: Transitions and rewards}
	\label{fig:durationRE}
\end{figure*}

\paragraph*{\textit{Action Execution}} Let $s$ be a state and $ad=\langle a,cost, \langle pre_1, \langle EF^1_1, p^1_1 \rangle,$ $..., \langle EF^1_m, p^1_m \rangle \rangle,..., \langle pre_n,$ $ \langle EF^n_1,$ $ p^n_1 \rangle,..., \langle EF^n_l, p^n_l \rangle \rangle\rangle \in \mathcal{AV}$ be the action description of $a\in \mathcal{A}$ in $\mathcal{D}_r$. The execution of $a$ in $s$ produces a state $r$ with a probability $p$ iff:
	\begin{itemize}
		\item a precondition $pre_i$ of the action description $ad$ is satisfied in $s$, i.e., $s\models pre_i$,
		\item one of the effects in $EF^i_j$ of $pre_i$ is $eff=\{l_1,...,l_n\}$,  
		\item the probability $p$ is $p^i_j$, 
		\item the state $r$ satisfies the following two conditions:
		\begin{itemize}
			\item 
			its base state is $\overline{s}$ after the update of the value of every state variable in which appears in $EF^i_j$ with the value specified in $EF^i_j$. Formally, this is represented as follows: $\overline{r}= (\overline{s}\setminus chg(\overline{s},EF^i_j)) \cup EF^i_j$ where $chg(\overline{s},EF^i_j)=\{x=v \,|\, x=v' \in EF^i_j,  x=v \in \overline{s} \}$, 
		
			\item its expanded state is $\dot{s}$ after the update of the state of every requirement according the state transition models shown in Fig.~(\ref{fig:unconditionalRE})(\ref{fig:conditional_deadline})(\ref{fig:durationRE}). Formally, $\dot{r}=\{upd_T(m,st,\overline{r}) \,|\, m=st \in \dot{s} \text{ and } type(m)=T\}$ where $upd_T(m,st,x)$ defines how the requirement $m$ of type $T$ should be updated when its current state is $st$ and the newly computed base state is $x$. This function is defined for every type of requirements according to its state transition model. For example, consider $PM$ requirements of the form $\textbf{ReqId }m \textbf{ maintain } S \textbf{ for } P$ $\textbf{ if } A \textbf{ unless } Z$ $\textbf{reward }r$, the definition of $upd_{PM}(m,st,x)$ is as follows:
	
		\begin{center}
			$\begin{displaystyle}
			upd_{PM}(m, st,x) = 
			\left.
			\begin{cases}
			m=A, & \text{if } st=I \text{ and } x \models A\\
			m=I, & \text{if } st=A \text{ and } x \models Z\\
			m=R(P), & \text{if } st=A \text{ and } x \models (S \, \& \, ! Z) \\	
			m=I, & \text{if } st=R(T) \text{ and } x\models Z\\	
			m=I, & \text{if } st=R(1) \\	
			m=R(T-1), & \text{if } st=R(T) \text{ and } x\not\models Z \text{ and } T\neq 1\\
			m=st, & \text{otherwise }
			\end{cases}
			\right. 
			\end{displaystyle}$
		\end{center}
		\end{itemize}
		\item Otherwise, if none of the action preconditions is true in $s$, then $\overline{r}=\overline{s}$, $\dot{r}=\{upd_T(m,st,\overline{r}) \,|\, (m=st) \in \dot{s} \text{ and } type(m)=T\}$ and $p=1$.

	\end{itemize}

\noindent Other functions are similarly defined to describe the update of the state of the other types of requirements as shown in the transitions part of Fig.~(\ref{fig:unconditionalRE})(\ref{fig:conditional_deadline})(\ref{fig:durationRE}). 
We define similarly the execution of an event $e$ in a state $s$ as follows.

\paragraph*{\textit{Event Execution}} Let $s$ be a state and $\langle e,\langle pre_1, op_1, \langle EF^1_1, p^1_1 \rangle,...,$ $\langle EF^1_m, p^1_m \rangle \rangle,..., \langle pre_n, op_n,$ $ \langle EF^n_1,$ $ p^n_1 \rangle,..., \langle EF^n_l, p^n_l \rangle \rangle\rangle \in \mathcal{ED}$ be the event description $ev$ of an event $e$ in $\mathcal{D}_r$. The execution of $e$ in $s$ produces a state $r$ with a probability $p$ iff:
\begin{itemize}
	\item a precondition $pre_i$ is satisfied in $s$, i.e., $s\models pre_i$,
	\item one of the effects in $EF^i_j$ of $pre_i$ is $eff=\{l_1,...,l_n\}$,  
	\item the probability $p$ is $p^i_j$, 
	\item the state $r$ satisfies the following two conditions:
	\begin{itemize}
		\item its base state is $\overline{s}$ after the update of the value of every state variable in which appears in $EF^i_j$ with the value specified in $EF^i_j$. Formally, this is represented as follows: $\overline{r}= (\overline{s}\setminus chg(\overline{s},EF^i_j)) \cup EF^i_j$ where $chg(\overline{s},EF^i_j)=\{x=v \,|\, x=v' \in EF^i_j,  x=v \in \overline{s} \}$, 
		
		\item its expanded state is $\dot{s}$ after the update of the state of every requirement according the state transition models shown in Fig.~(\ref{fig:unconditionalRE})(\ref{fig:conditional_deadline})(\ref{fig:durationRE}). Formally, $\dot{r}=\{upd^e_T(m,st,\overline{r}) \,|\, m=st \in \dot{s} \text{ and } type(m)=T\}$ where $upd^e_T(m,st,x)$ defines how the requirement $m$ of type $T$ should be updated when its current state is $st$ and the newly computed base state is $x$ due to an event occurrence. The function $upd^e_T(m,st,x)$ is defined similarly to $upd_T(m,st,x)$ with the exception that events do not cause time-related transitions in the requirements' state machines since they occur concurrently with actions. For example, consider $PM$ requirements of the form $\textbf{ReqId }m \textbf{ maintain } S \textbf{ for } P \textbf{ if }$ $A \textbf{ unless } Z \textbf{ reward }r$, the definition of $upd^e_{PM}(m,st,x)$ is as follows:
		
		\begin{center}
			$\begin{displaystyle}
			upd^e_{PM}(m, st,X) = 
			\left.
			\begin{cases}
			m=A, & \text{if } st=I \text{ and } x \models A\\
			m=I, & \text{if } st=A \text{ and } x \models Z\\
			m=R(P), & \text{if } st=A \text{ and } x \models (S \, \& \, ! Z) \\	
			m=I, & \text{if } st=R(T) \text{ and } x\models Z\\	
			m=I, & \text{if } st=R(1) \\	
			m=st, & \text{otherwise }
			\end{cases}
			\right. 
			\end{displaystyle}$
		\end{center} 
	\end{itemize}
	\item Otherwise, if none of the event preconditions is true in $s$, then $\overline{r}=\overline{s}$, $\dot{r}=\{upd^e_T(m,st,\overline{r}) \,|\, (m=st) \in \dot{s} \text{ and } type(m)=T\}$ and $p=1$.

\end{itemize}

\comments{ $\mathcal{L}^{\mathcal{D}_r}_{\mathcal{RQ}}$ be the sets of state atoms and requirement atoms generated from $\mathcal{D}_r$ respectively.

An assignment of a value $v_i\in dom(x_i)$ to a state variable $x_i$ is called a {\it state atom} over $x_i$. The set of state atoms $\mathcal{L}^{\mathcal{D}_r}_{\mathcal{AV}}=\{x_i$=$v_j \,|\, x_i\in \mathcal{ID}_s \textsl{ and } v_i\in dom(x_i)\}$ is called {\it the set of state atoms generated from $\mathcal{D}_r$}.

Let $\mathcal{V}^{\mathcal{D}_r}=\mathcal{AV}\cup \mathcal{L}^{\mathcal{D}_r}_{\mathcal{AV}}$ and $\mathcal{L}^{\mathcal{D}_r}_{\mathcal{RQ}}$ be the sets of state atoms and requirement atoms generated from $\mathcal{D}_r$ respectively. 

The set of atoms generated from $r$ would be $\{m=I, m=R\}$. More generally, let $r\in \mathcal{RQ}$ be a requirement having one of the fourteen types introduced in Sec.\ref{sec:modeling_requirements}.

For simplicity, when a state variable $x_i$ is binary.

We use $var(l)$ and $val(l)$ to denote the state variable and value of a literal $l$ respectively.

is a set of {\it state variables}

the set of states generated from $\mathcal{X}$ is the set $\mathcal{S} = \{\bigcup_{i=1}^n \{x_i=v_i\} \,|\, v_i\in dom(x_i)\}$. 
\noindent Intuitively, the previous definition means that every state $s\in \mathcal{S}$ is a set of atoms such that every variable $x\in \mathcal{X}$ is assigned a value from its domain.

\paragraph*{Construction of States}
The states $\mathcal{S}$ correspond to all possible configurations of the system and the environment. A state represents a specific {\it configuration}. A configuration is an assignment of: (1) every variable in $\mathcal{SV}$ a value from its domain, (2) every requirement $r\in \mathcal{R}$ a value representing one state from its corresponding state machine (see Fig.~\ref{fig:statetransitionsrewards}). 
Thus, every state may be viewed as consisting of two parts: (1) a base state consisting of a specific configuration of state variables, and (2) an expanded state consisting of a specific configuration of requirements. 

For example, consider a domain model $\mathcal{D}_r$ comprising of two boolean variables $x$ and $y$ and one requirement $m$ of type $CA$. The states $\mathcal{S}$ constructed on the basis of $\mathcal{D}_r$ represent all possible configurations of its state variables and requirements. Thus, $\mathcal{S}$ includes the eight states in Fig.~\ref{fig:example-states}.

\begin{figure}[t]
	\centering
	\includegraphics[width=0.8\linewidth]{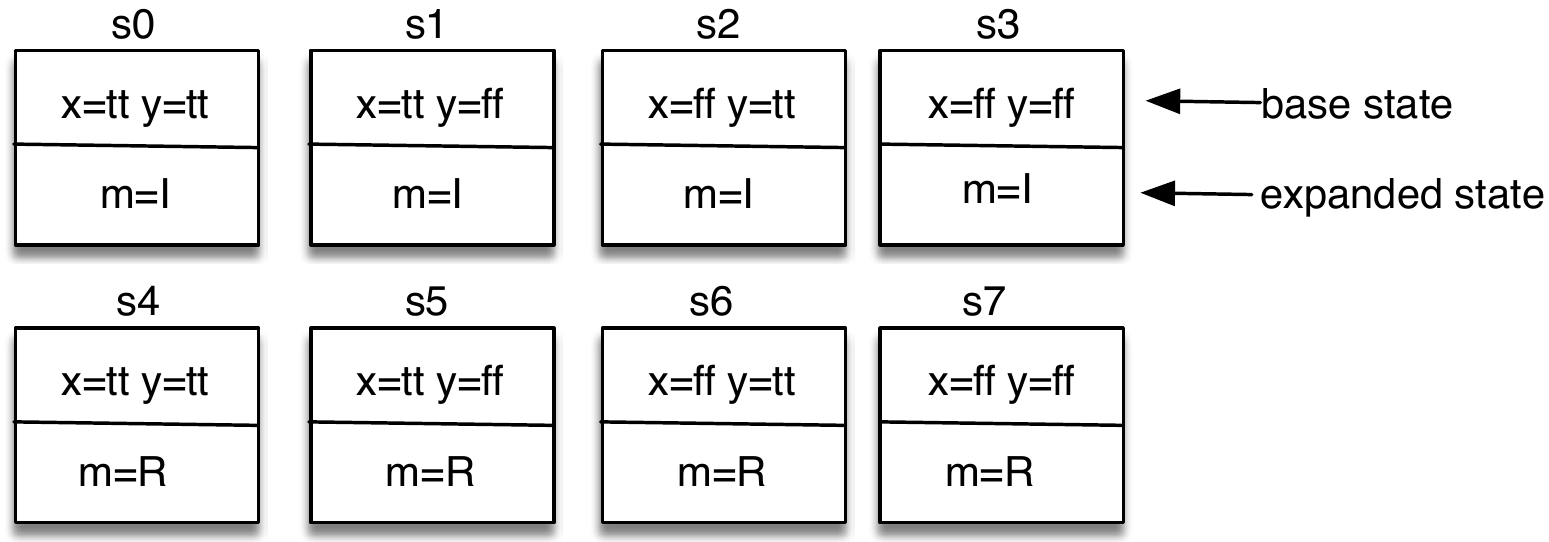}
	\caption{Constructed States}
	\label{fig:example-states}
\end{figure}

}

\paragraph*{\textit{Event Occurrence Vector}} Let $s$ be a state and $\langle e,\langle pre_1, op_1, \langle EF^1_1, p^1_1 \rangle,$ $..., \langle EF^1_m, p^1_m \rangle \rangle,$ $...,$ $\langle pre_n, op_n,$ $ \langle EF^n_1,$ $ p^n_1 \rangle,..., \langle EF^n_l, p^n_l \rangle \rangle\rangle \in \mathcal{ED}$ be an event description $ed$ of an event $e$ in $\mathcal{D}_r$. 
	The occurrence vector of $e$ is a vector $O_e$ of length $|\mathcal{S}|$ whose entries are defined as follows:

\begin{center}
	$\begin{displaystyle}
	O_e(s) = 
	\left.
	\begin{cases}
	op_i & \text{if } s \models pre_i\\	
	0 & \text{otherwise }
	\end{cases}
	\right. 
	\end{displaystyle}$
\end{center}

\paragraph*{\textit{Explicit Action Transition Matrix}} Let $a\in \mathcal{A}$ be an action and $\mathcal{S}$ be the set states. The {\it explicit transition matrix} of $a$, denoted $Pr_a$, is a $|\mathcal{S}|\times |\mathcal{S}|$ matrix. If the execution of $a$ in a state $s\in \mathcal{S}$ produces the state $r\in \mathcal{S}$ with a probability $p$, then $Pr_a(s,r)=p$.

\paragraph*{\textit{Explicit Event Transition Matrix}} Let $e$ be an event and $\mathcal{S}$ be the set states. The {\it explicit transition matrix} of $e$, denoted $Pr_e$, is a $|\mathcal{S}|\times |\mathcal{S}|$ matrix. If the execution of $e$ in a state $s\in \mathcal{S}$ produces the state $r\in \mathcal{S}$ with a probability $p$, then $Pr_e(s,r)=p$.

\paragraph*{\textit{Effective Events Transition Matrix}} Let $Pr_{e_1},...,Pr_{e_n}$ be the explicit event transition matrices of events in $\mathcal{D}_r$ and $O_{e_1},...,O_{e_n}$ be their corresponding occurrence vectors. Let $E,\;E'$ be the diagonal matrices with entries $E_{kk}=O_e(s_k)$ and $E'_{kk}=1-O_e(s_k)$ respectively. The {\it effective transition matrix} of an event $e_i \in \{e_1,...,e_n\}$, denoted $\hat{TM}_e$, is computed as follows:
	\comments{
		\begin{equation}
		\hat{Pr}_e(s_i,s_j)=O_{e}(s_i)Pr_e(s_i,s_j) + 
		\begin{cases*}
		1-O_{e}(s_i) & :  i=j \\
		0        & :  $i \neq j$
		\end{cases*}
		\end{equation}
	}
	$$\hat{Pr}_{e_i}=(E\times Pr_{e_i}) + E'$$
	Given the effective transition matrices of the events $e_1,..,e_n$, the effective events transition matrix, denoted $TM_{ev}$ is computed as follows:
	$$Pr_{ev}= \hat{Pr}_{e_1}\times...\times\hat{Pr}_{e_n} $$

\begin{definition}[Action Transition Matrix] Let $Pr_{a_1},...,Pr_{a_n}$ be the explicit action transition matrices of actions in $\mathcal{D}_r$ and $Pr_{ev}$ be the effective events transition matrix. The ({\it implicit-event}) {\it action transition matrix} of an action $a_i \in \{a_1,...,a_n\}$, denoted $\hat{Pr}_{a_i}$, is computed as follows:
		$$\hat{Pr_{a_i}}= Pr_{a_i} \times Pr_{ev} $$
\end{definition}


\begin{definition}[Action Reward matrix] Let $a\in \mathcal{A}$ be an action and $\mathcal{S}$ be the set states. The {\it action reward matrix} of $a$, denoted $R_a$, is a $|\mathcal{S}|\times |\mathcal{S}|$ matrix such that if $s_i$ and $s_j$ are states in $\mathcal{S}$, then $R_a(s_i,s_j)$ represents the rewards that are obtained on the transition from the state $s_i$ to $s_j$ minus the cost of the action $a$. This is expressed as follows: $R_a(s_i,s_j)= (\sum^{|RS|} RS) - cost(a)$ where $RS=\{   rew_T(m,s_i,s_j)  \, |\, type(m)=T \textbf{ and } (m=st_i)\in \dot{s_i} \text{ and } (m=st_f)\in \dot{s_j} \}$. The function $rew_T(m,s_i,s_j)$ is defined for every requirement according to its type as shown in the rewards part of Fig.~(\ref{fig:unconditionalRE})(\ref{fig:conditional_deadline})(\ref{fig:durationRE}).
For example, consider a $PM$ requirements of the form $\textbf{ReqId }m \textbf{ maintain } S \textbf{ for } P \textbf{ if } A \textbf{ unless } Z$ $\textbf{reward }r$, the definition of $rew_{PM}(m,s_i,s_j)$ is as follows:
\begin{center}
$\begin{displaystyle}
rew_{PM}(m,s_i, s_j) = 
\left.
\begin{cases}
r, & \text{if } s_i\models (S \& ((m$=$R(X)) || ... || (m$=$R(1))) \text{ and }\\
& \hspace{1cm} s_j\models (S \& (m$=$R(X) || ... || (m=R(1)))\\
0, & \text{otherwise }
\end{cases}
\right. 
\end{displaystyle}$
\end{center}
\end{definition}

\noindent Reward functions for the other types of requirements are similarly defined.
\begin{definition}[The discount factor] The discount factor $\gamma$ is a value between zero and one, i.e., $0<\gamma<1$.
\end{definition}
\noindent The discount factor ensures the convergence of the infinite reward series when computing the total expected rewards. It determines how far into the future the satisfaction of requirements affects the computation of optimal strategies. For example, if $\gamma$ be 0.98 and $r$ is the reward defined for requirement $m$. In this case, the actual reward values obtained if this requirement is satisfied after 50, 100 and 150 time steps are 0.364r, 0.1326r and 0.0482r respectively. The discount factor is therefore chosen according to the requirements of the application domain. 

